\documentclass[aps,prl,reprint,groupedaddress,showpacs,amsmath,amssymb,twocolumn,floatfix]{revtex4-1}
\usepackage{graphicx}
\usepackage{bm}
\usepackage{color}
\usepackage[normalem]{ulem}
\usepackage{cleveref}
\usepackage{amsmath}

\newcommand{\head}[1]{
\vspace{\baselineskip}
\noindent\textbf{#1 \medskip}\\
\noindent
}

\begin{document}

\title{Quantitative measurement of viscosity in two-dimensional electron fluids}

\author{Yihang Zeng$^{1,5}$}
\author{Haoyu Guo$^{2,4}$}
\author{Olivia M. Ghosh$^{1}$}
\author{Kenji Watanabe$^{3}$}
\author{Takashi Taniguchi$^{3}$}
\author{Leonid S. Levitov$^{4}$}
\author{Cory R. Dean$^{1}$$^{*}$}

\affiliation{$^{1}$Department of Physics, Columbia University, New York, NY, USA}
\affiliation{$^{2}$Laboratory of Atomic and Solid State Physics, Cornell University, Ithaca, NY, USA}
\affiliation{$^{3}$National Institute for Materials Science, Tsukuba, Japan}
\affiliation{$^{4}$Department of Physics, Massachusetts Institute of Technology, Cambridge, Massachusetts 02139, USA}
\affiliation{$^{5}$Present address: Department of Physics, Purdue University, West Lafayette, Indiana, USA}

\date{\today}



\maketitle

\textbf{Electron hydrodynamics is an emerging framework that describes dynamics of interacting electron systems as conventional fluids. While evidence for hydrodynamic-like transport is reported in a variety of two-dimensional materials, precise quantitative measurement of the core parameter,  electron viscosity, remains challenging.  In this work, we demonstrate that magnetoresistance in Corbino-shaped graphene devices offers a simultaneous Ohmmeter/viscosometer, allowing us to disentangle the individual Ohmic and viscous contributions to the transport response, even in the mixed flow regime.  Most surprising, we find that in both monolayer and bilayer graphene, the effective electron-electron scattering rate scales linearly with temperature, at odds with the expected $T$-squared dependence expected from conventional Fermi liquid theory, but consistent with a recently identified tomographic flow regime, which was theoretically conjectured to be generic for two-dimensional charged fluids.}

\begin{figure*}
\includegraphics[width=0.95\linewidth]{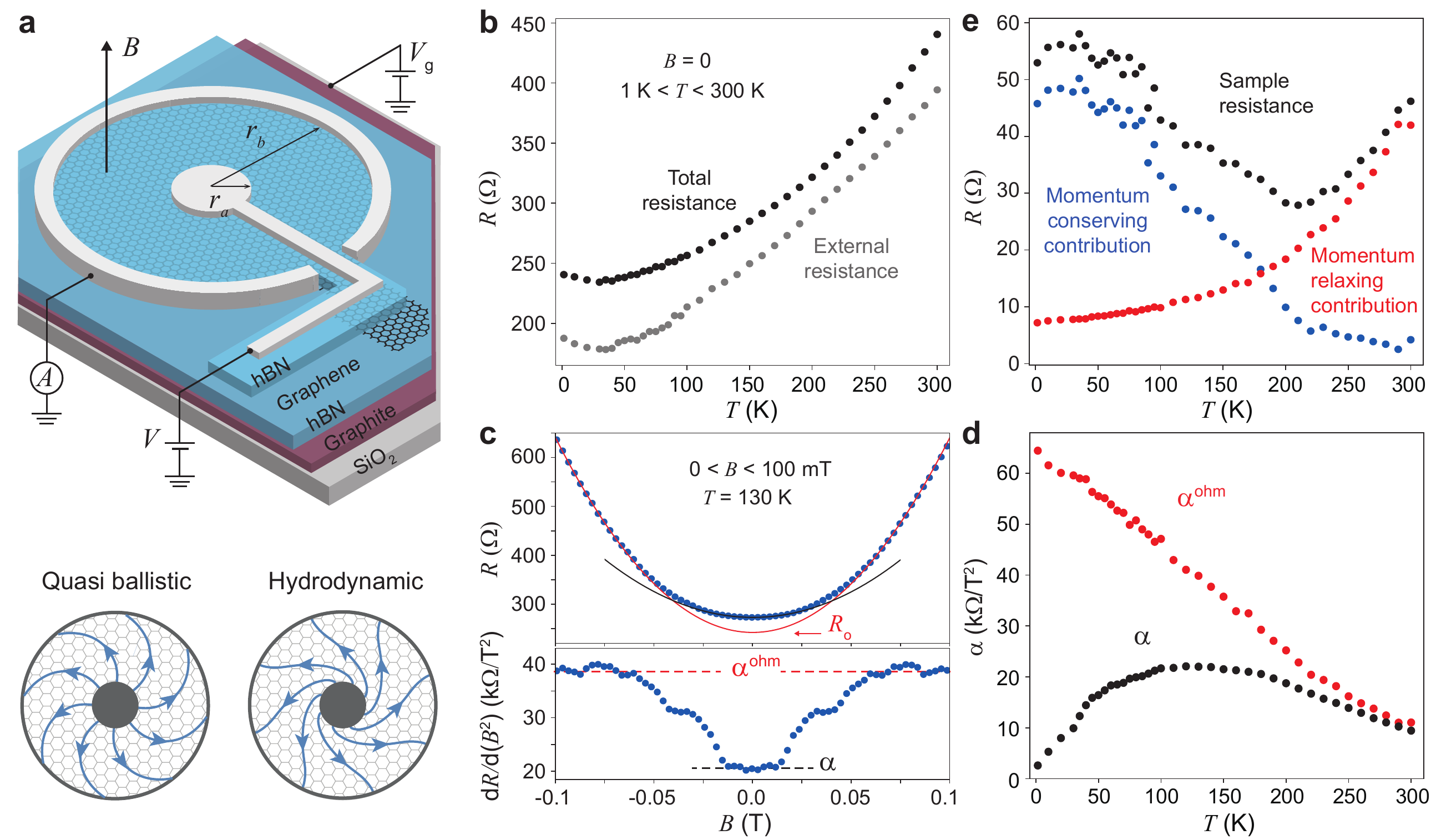}
\caption{\label{line} {\bf{Magnetoresistance of graphene in the corbino disk geometry.}} \textbf{a},Schematic of the corbino disk geometry and measurement configuration. A small opening of width $\approx r_a/4$ is left out on the top hBN layer at the outer circumference in order for the inner electrode to pass through. Bottom schematics depict current flow lines under small perpendicular magnetic field while in the quasi ballistic (left) and hydrodynamic (right) transport regime. \textbf{b} Zero-magnetic-field resistance $R(B=0)$ and external resistance $R^{Ext}$ as a function of temperature. The external resistance consists of both the metal-graphene contact resistance and line resistance. \textbf{c},Representative magnetoresistance and its derivative against $B^2$ at $T=$~130~K. The derivative is obtained by fitting a quadratic function to symmetrized magnetoresistance in a sliding window of 12~mT wide. The plateau at low field and high field show two quadratic regimes with curvature $\alpha$ and $\alpha^{ohm}$ respectively. $R_o$ denotes the intercept of the high-field quadratic function. The determination of fitting range is discussed in SI. \textbf{d}, $\alpha$ and $\alpha^{ohm}$ versus temperature. Results in b-d are taken at $n=0.2\times10^{12}$cm$^{-2}$ in a device with $r_a=1\mu$m, $r_b=4\mu$m. \textbf{e},Sample resistance as a function of temperature. Black circles mark total sample resistance. Red circles show resistance from momentum-relaxing scattering. While blue circles represent the difference between black and red circles, reflect the resistance contributed by momentum-conserving scattering.}
\end{figure*} 


Strongly correlated electron systems in which many-body interactions drive novel ground states remain one of the most fascinating, yet difficult to understand, problems in physics\cite{alexandradinata2022futurecorrelatedelectronproblem}. 
A particular challenge is relating the macroscale behaviour typically measured in experiment, such as transport response, with the underlying microscopic interactions and scattering processes relevant to theory. 
 Electron hydrodynamics has emerged as a framework to to fill this gap by identifying that strong interparticle scattering can cause electron systems to behave like a conventional viscous fluid\cite{Molenkamp1995prb,damle1997,Davison2014,forcella2014,Levitov2016}. 
  The viscous flow response is parameterized by the electron viscosity and can therefore, in principle, be used as a measure of electron-electron scattering\cite{gurzhi1962nonmonotonic, Gurzhi_1968, muller2009perfectfluid}.
Quantifying this parameter in experiment is especially relevant in two-dimensional (2D) electron systems where theory predicts different families of excitations distinguished by quasiparticle lifetime\cite{ledwith2019,ledwith2019hierarchy,Hofmann2023longlifetimes,Hofmann2022collectivemodes,Kryhin2023,nilsson2024nonequilibriumrelaxationoddeveneffect,hong2020superscreeningretroreflectedholebackflow}. These long-lived excitations are hard to detect by ordinary spectroscopy techniques\cite{Eisenstein2007quantumlifetime,Murphy1995lifetimetunneling}, however they are expected to  dominate the hydrodynamic response\cite{ledwith2019,Kryhin2023}.

Evidence for electron hydrodynamics has been identified in a variety of 2D electron systems\cite{lucas2018reviewhydrodynamics, LevchenkoTransportPropertiesOf2020}. However, precise quantitative comparison between experiment and theory remains limited. 
This is the result of two challenges.  First, electron-electron ($e$-$e$) scattering is a momentum conserving interaction that does not directly yield a measurable dissipation. Instead energy is lost through momentum-relaxing boundary scattering.  Dissipation in the hydrodynamic regime is therefore determined by device geometry and boundary conditions, and only indirectly related to  $e$-$e$ scattering (i.e.viscosity). Second, for a material to demonstrate purely hydrodynamic flow requires a precise combination of properties, namely long mean free path for momentum-relaxing scattering (including electron-impurity, electron-phonon and Umklapp scattering) $l_{ee}\ll l_{mr}$, and device dimensions larger than $l_{ee}$ but smaller than $l_{mr}$. In real devices this limit is typically only approximately reached so that both hydrodynamic and Ohmic processes contribute to the observed response, and disambiguating their relative contributions is difficult and often imprecise.  Experimental progress has been realized through device-size scaling measurements \cite{Molenkamp1994prb,Molenkamp1995prb, Moll2016,kumar2017superballistic,Ginzburg2023}, non-local transport\cite{Bandurin2016,Berdyugin2019,Kim2020viscosityscreening}, thermal transport\cite{talanov2024viscousdissipationthermal}, and direct imaging of flow profiles\cite{Ku2020NVhydro,Kumar2022LS,sulpizio2019, Jenkins2022NVimaginghydrographene, Aharon-Steinberg2022observationvortice,Vool2021imagingviscous,Braem2018scanninggateviscous}.  In most cases hydrodynamic parameters are deduced only in the regime where it is presumed that Ohmic scattering is minimal, restricting the available phase space and limiting the ability to deduce scaling behaviours (e.g. density and temperature dependence).

In this work we demonstrate a new approach to measuring electron viscosity in 2D systems based on  magnetotransport in a Corbino geometry.  The Corbino structure measures current flow betweeen concentric rings and therefore has no boundary except at the contacts.  Recent theoretical\cite{levchenko2020, LevchenkoTransportPropertiesOf2020, shavit2019_corbino, holder2019} and experimental\cite{Kumar2022LS} work has demonstrated that the potential distribution and current-flow profiles through the Corbino disc are dramatically different in the Ohmic versus viscous regimes. We show that under mangetic field, the magnetoresistances associated with each regime (Fig. 1a) can be utilized to quantify the resistivity and viscosity, respectively\cite{shavit2019_corbino}. 
Moreover, we find that it is possible to disentangle the two contributions in the mixed flow regime, enabling a powerful way to isolate the viscous parameters over a wider phase space than previously possible.

We apply this technique to the study of hydrodynamic flow in monolayer (MLG) and bilayer (BLG) graphene, and make the surprising discovery that in both cases, in the Fermi liquid regime,  the electron kinetic viscosity scales with $1/T$ and is density-independent. This violates conventional Fermi liquid behaviour but finds consistency with a newly-identified tomographic flow regime, theoretically predicted for 2D materials\cite{ledwith2019,Hofmann2022collectivemodes,Kryhin2023}.  We also demonstrate continuous measurement of viscosity through the crossover from the Fermi liquid regime at high density to Dirac fluid regime at low density.  We show for the first time that the MLG and BLG viscosity transitions to a $T^{2}$, and $T^{1}$ dependence, respectively,  in the Dirac fluid regime, consistent with theoretical predicitons\cite{muller2009perfectfluid, yudhistira2023nonmonotonic}, and confirm that in the crossover the system behaves according to a two-fluid model with $e$-$e$ and $e$-$h$ contributing to hydrodynamic electrical transport in two parallel channels\cite{yudhistira2023nonmonotonic}.

\head{Magnetoresistance in a corbino disk}
Fig. 1a shows a cartoon schematic of our device structure. Graphene is encapsulated between boron nitride layers. A corbino geometry with concentric inner and outer ring electrodes is then defined lithographically using an edge contact geometry\cite{Zeng2019}. A local graphite bottom gate is used to tune the carrier density (see methods for more details). We investigate MLG and BLG devices, with outer radius, $r_{b}$, ranging in size from 1.75~$\mu$m to 4.5~$\mu$m, and inner radius $r_{a}=r_{b}/4$, maintaining  a constant aspect ratio across all devices.


Fig. 1b plots the two-terminal resistance as a function of temperature for a device with $r_b = 4 \mu$m, measured at electron density $n=0.2\times10^{12}$cm$^{-2}$. 
Interpreting the Corbino response is challenging since our measurement spans a mixed flow regime that includes both viscous and Ohmic contriubtions in addition to contact resistance, and external line resistance, all of which may vary with temperature and density.  In the following we demonstrate how analysis of the magnetoresitance allows us to disentangle these individual contributions.

The Corbino magnetoresistance includes a mixture of longitudinal and Hall components giving a quadratic field dependence, $R(B)=R_{o}+\alpha B^2$.  This holds true in the presence of both momentum-relaxing and momentum-conserving scattering\cite{Weiss1954corbino,Blood1971corbino,wieder1969corbino,shavit2019_corbino,LevchenkoTransportPropertiesOf2020}, so long as the individual scattering terms and contact resistances have a negligible field dependence. $R_{o}$ is the zero-field resistance, including all dissipative contributions. The quadratic coefficient, $\alpha$, arises from a combination of the Hall voltage, scattering mechanisms, and boundary conditions, and does not include external resistance contributions, allowing us to isolate dissipation across the channel only \cite{Weiss1954corbino,wieder1969corbino,Blood1971corbino,shavit2019_corbino,LevchenkoTransportPropertiesOf2020}. The coefficient can be solved for in the mixed flow regime using a combined Stokes-Ohm equation in magnetic field (see methods) to give an analytic expression that depends on both the ohmic resistivity, $\rho_{o}$, and shear viscosity, $\eta$\cite{LevchenkoTransportPropertiesOf2020}:





\begin{equation}
    \alpha=\frac{\ln\gamma}{2\pi\rho_{o} (ne)^2}[1-f(\gamma,r_{b}/D_{\nu})]
\end{equation}

\noindent
where $f(\gamma,r_{b}/D_{\nu})$  is a dimensionless function defined in terms of modified Bessel functions (see methods), $\gamma=r_{b}/r_{a}$ is the device aspect ratio,  and $D_{\nu}=\sqrt{{\eta}/{\rho_{o}}(ne)^2}$ is the Gurzhi length. We note that when $D_{\nu}$ is much larger than the device size, $D_{\nu} \gg r_b$, then Eq. 1 reduces to a purely hydrodynamic form that describes viscous flow\cite{shavit2019_corbino}

\begin{equation}
    \alpha^{visc.}=\frac{r_b^2}{16\pi\eta}[1-\frac{1}{\gamma^2}-\frac{4\ln^2\gamma}{(\gamma^2-1)}].
\end{equation}

\noindent
In the opposite limit,  $D_{\nu} \ll r_a$,  Eq. 1 reduces to a purely Ohmic form\cite{Weiss1954corbino}, 

\begin{equation}    
\alpha^{Ohm}=\frac{\ln{\gamma} }{2\pi\rho_{o}(ne)^2}.
\end{equation}

Fig. 1c upper panel shows the magnetoresistance measured at $T=130$~K for the same carrier density as in Fig. 1b.  The resistance shows the expected $B^{2}$  dependence, but with two apparent regimes. This is seen more clearly by plotting the ${dR}/{d(B^2)}$ derivative of the magnetoresistance (Fig. 1c, lower panel), which shows two distinct plateaus. By fitting two separate parabolas to the magnetoresistance we extract two curvatures corresponding to the ``low'' and ``high'' field regimes, respectively  
 (black and red curves in Fig. 1b -- see the determination of fitting range in SI).

We assume $\rho_{o}$ is constant in magnetic field, however the viscosity is modified with $B$
according to $\eta(B) = {\eta}/({1+4l_{ee}^{2}/r_c^2})$, where $r_{c}=\hbar\sqrt{\pi n}/eB$ is the cyclotron radius\cite{Alekseev2016,holder2019}.  As $B$ increases, $r_{c}$ decreases and eventually $\eta\rightarrow0$, fully suppressing the viscous contribution and yielding a predominantly Ohmic response.  We therefore interpret the two curvatures as follows:  the low field curvature, labelled  $\alpha$, describes the mixed flow regime and thus depends on both $\eta$ and $\rho_{o}$ (Eq. 1); the high field curvature, labelled $\alpha^{Ohm}$ reflects the transition to purely Ohmic response, and thus depends on $\rho_{o}$ only (Eq. 3). 

Fig. 1d shows the temperature dependence of the two quadratic coefficients. The high field curvature, $\alpha^{Ohm}$, decreases monotonically, as expected from Eq. 3 and assuming $d\rho_{o}/dT>0$ (typical for graphene at these densities). The low field curvature,  shows a non-monotonic trend, first increasing then decreasing.  This is consistent with the transport response transitioning from being more viscous-like at low temperature to Ohmic-like at high temperature.  The crossover results from the fact that while $\alpha^{visc}\sim1/\eta$ (Eq. 2) and $\alpha^{Ohm}\sim 1/\rho_{o}$ (Eq. 3), $\eta$ decreases with $T$ whereas $\rho_{o}$ increases with $T$.  

The ability to isolate the Ohmic magnetoresistance at high field provides a means to measure the Ohmic scattering term, independent of any zero-field contact or hydrodynamic contributions and without any other free parameters. Once we determine $\rho_{o}$, we can then use our measurement of low-field $\alpha$ along with Eq. 1 to extract $\eta$. This same process can be performed at any density and temperature, allowing us to quantify the individual Ohmic and viscous flow-related parameters throughout the mixed-flow regime.

\begin{figure*}
\includegraphics[width= 0.95\linewidth]{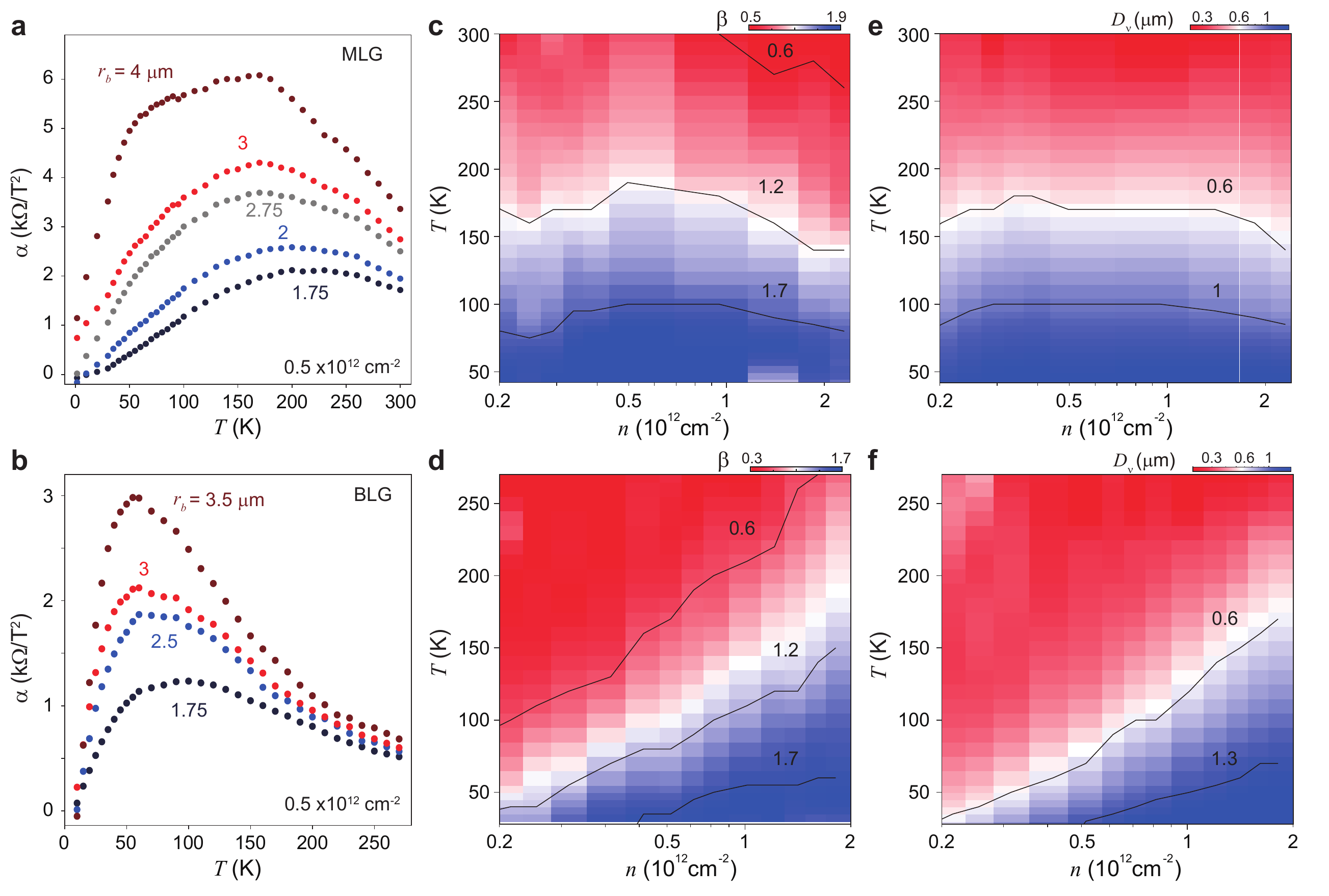}
\caption{\label{pattern}{\bf{Scale-dependent magnetotransport and viscous-ohmic crossover}} \textbf{a,b},$\alpha$ vs $T$ in five MLG corbino devices (a) and four BLG corbino devices (b) with the same aspect ratio $\gamma=4$ but different radii. \textbf{c,d},The scaling power $\beta$ which is extracted by fitting $\alpha$ vs $r_b$ for MLG (c) and BLG (d) devices as a function of $n$ and $T$ (see SI for examples on fitting). Contours are marked by dashed lines. \textbf{e,f}, Measured Gurzhi length $D_{\nu}$ versus $n$ and $T$ for MLG (e) and BLG (f).}
\end{figure*}

First, we focus on the Ohmic resistance. The red data points in Fig. 1e show a plot of $\frac{\ln\gamma}{2 \pi}\rho_{o}(T)$ measured at the same density as Fig. 1b,d,e.  The temperature dependence of $\rho_{o}(T)$ closely resembles results from four-terminal Hall bar measurement (see SI) where the linear and exponential terms arise from acoustic and optical phonon scattering, respectively, and any impurity scattering contribution remains temperature-independent over this range\cite{Dean10,chen2008graphene_scattering}. Measurement over a wide density range for both MLG and BLG show similar results (see SI). Fitting the linear-T regime gives an acoustic-phonon deformation potential $D_A=18 \pm 1$~eV in quantitative agreement with prior measurements, which we take as confirmation that we are faithfully determining $\rho_{o}$ with our analysis. 


The $B=0$ intercept of the parabolic fit to the $\alpha^{Ohm}$ regime, $R_{o}$, includes the zero field Ohmic and external resistances, $R_{o}=\frac{\ln\gamma}{2 \pi}\rho_{o}+R^{Ext}$.
 Using the value of $\rho_{o}$ determined from the curvature together with $R_{o}$ therefore allows us to obtain a value for $R^{ext}$. Fig. 1b shows a plot of $R^{ext}(T)$ (grey circles) compared with the total two-terminal resistance, $R^{tot}(T)$ (black circles).  The two curves show a similar trend and magnitude, indicating a significant proportion of the overall response is dominated by the external resistances.  Nonetheless, the sample-only resistance, $R^{sample}$ can be obtained by subtracting these two curves $R^{sample}=R^{tot}-R^{ext}$. The black circles in Fig. 1e shows a plot of $R^{sample}(T)$. We observe a non-monontonic dependence, first decreasing and then increasing with temperature. This behaviour is a manifestation of the Gurzhi effect\cite{gurzhi1962nonmonotonic,Gurzhi_1968}, which predicts the observed $dR/dT$ sign change as the system evolves from momentum-conserving (viscous flow) to momentum-relaxing (Ohmic flow) dominated scattering with increasing temperature\cite{Molenkamp1994prb,Molenkamp1995prb, Moll2016}. Finally, we isolate the viscous-flow contribution to the sample resistance according to the Matthiessen's rule,  $R^{visc}=R^{sample}-R^{Ohmic}$, shown as blue circles in Fig. 1e.  $R^{visc}$ decreases monotonically with temperature, as expected for hydrodynamic transport \cite{shavit2019_corbino,Kumar2022LS}.

We can further confirm our interpretation of the viscous-Ohmic crossover with increasing temperature by looking at how the  magnetoresistance scales with device size.  In the purely viscous flow regime, $\alpha$ varies with the outer radius squared (Eq. 2), whereas in the purely Ohmic limit $\alpha$ depends on the aspect ratio only (Eq. 3), independent of the size.  Fig. 2a (2b) shows $\alpha$ versus temperature measured in MLG (BLG) devices with size ranging from $r_b=1.75 \mu$m to $4 \mu$m ($r_b=1.75 \mu$m to $3.5 \mu$m), but fixed $\gamma=4$. The different radius Corbino devices where fabricated from a single heterostructure to ensure identical scale-independent properties (see SI). 
To quantify the size dependence we calculate the scaling power by assuming $\alpha\sim r_{b}^{\beta}$ and then determine $\beta$ from a linear fit on a log-log plot (see SI for examples of this fitting). Fig. 2b,c plots $\beta$ versus density and temperature for MLG and BLG, respectively. In both cases, $\beta\sim 2$ at low $T$ and then decreases monotonically with increasing $T$.  This is consisent with a predominantly hydrodynamic flow at low $T$, but with Ohmic contributions increasingly participating with increasing $T$.   $\beta$ never reaches zero, suggesting that in the density range considered,  non-negligible viscous contributions remain present up to at least $T\approx300$~K.  Additionally we observe that $\beta$ contour lines are overall density indpendent for MLG, but density dependent in BLG reflecting differences in the $\eta(n)$ and $\rho_{o}(n)$ for the two systems.  In Fig. 2e,f we plot the Gurzhi length versus $T$ and $n$, using the magnetoresistance curvature to determine $\rho_{o}$ and $\eta$, as described above.  The countour lines of  $D_{\nu}$ exhibit a similar trend as the corresponding $\beta$, confirming that the viscous-Ohmic crossover coincides with a reduction in $D_{\nu}$, as expected.

\begin{figure*}
\includegraphics[width=0.95\linewidth]{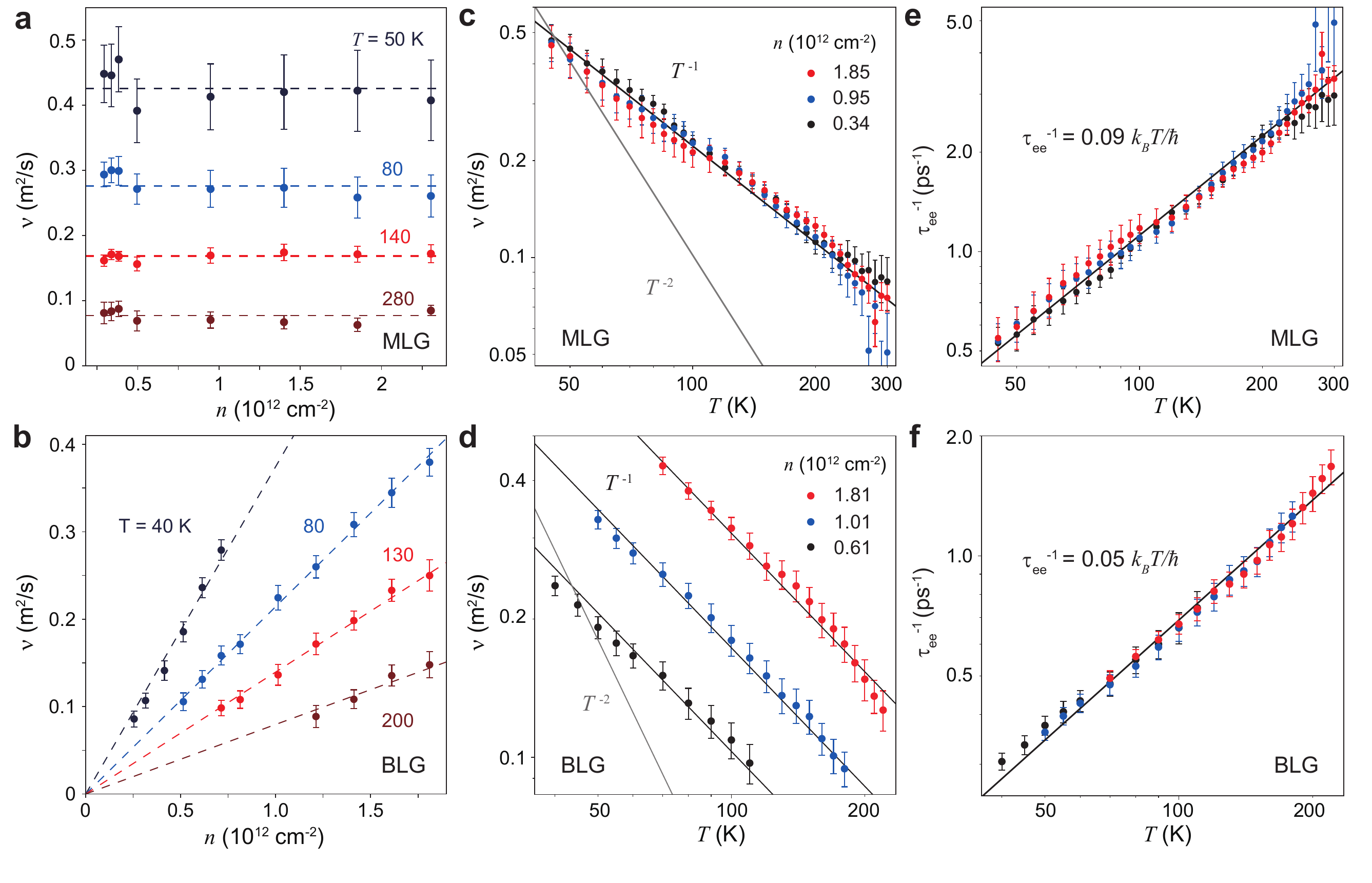}
\caption{\label{datamap} {\bf{Kinematic viscosity and effective electron-electron scattering rate in doped graphene.}} \textbf{a,b} Kinematic viscosity $\nu$ versus carrier density at different temperatures for MLG (a) and BLG (b). \textbf{c,d},Kinematic viscosity versus temperature for heavily doped MLG (c) and BLG (d). Black solid line denotes fit to $T^{-1}$. The dependence $T^{-2}$ is shown in grey as a comparison. \textbf{e,f},Electron-electron scattering rate calculated from $\nu$ in (c),(d),respectively, using $\nu=v_F^2 \tau_{ee}/4$. Data present in this figure is taken where $E_F/k_BT>2.5$.}
\end{figure*}

\begin{figure*}
\includegraphics[width=0.95\linewidth]{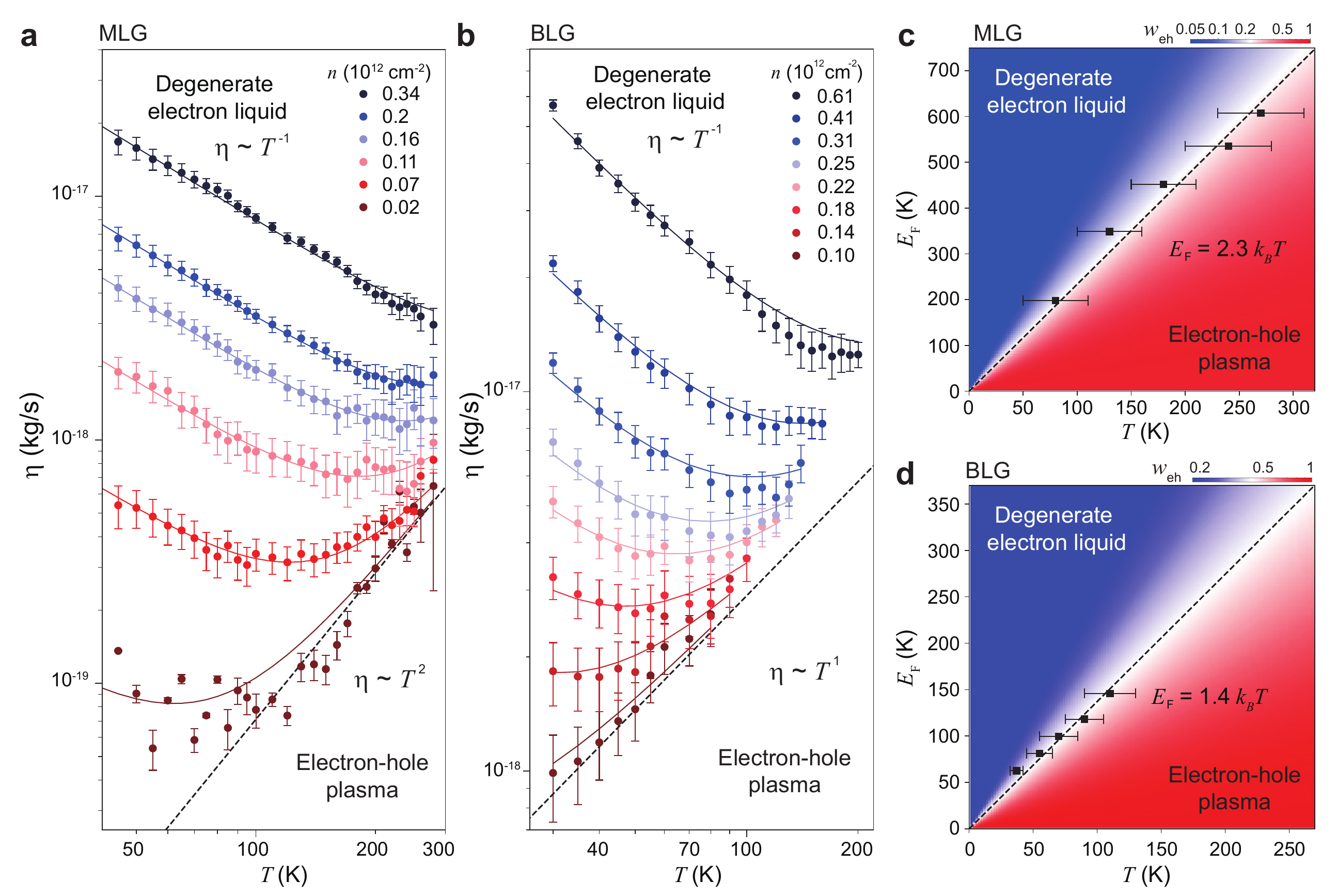}
\caption{\label{datamap} {\bf{Crossover between Fermi liquid and electron-hole plasma.}} \textbf{a},Shear viscosity $\eta$ versus temperature at different carrier densities in MLG. Black dashed line denotes the $\eta\sim T^2$ dependence. Solid lines are fitting to data using Eq. 4 (see SI for fitting parameters). \textbf{b},$\eta$ versus temperature at different carrier densities in BLG. Black (gray) dashed line denotes the $\eta\sim T$ ($\sim T^2$) dependence. Solid lines are fits to data. \textbf{c,d} Percentage of viscosity from electron-hole plasma in total viscosity $w_{eh} = \eta_{eh}/\eta$ for MLG (c) and BLG (d), respectively. $w_{eh} = \eta_{eh}/\eta$ is reconstructed from fitted functions (see SI) and Eq. 4.
Scattered data points mark the crossover temperature defined as where $\eta$ is minimized.}
\end{figure*} 


\head{Anomalous Fermi liquid behaviour}
Next we examine how the kinematic viscosity $\nu=\eta/nm$  evolves with temperature and carrier density, as $\nu$ directly captures momentum-diffusion in the equations of motion\cite{yudhistira2023nonmonotonic}.
For MLG, we use the mass definition $m=\hbar\sqrt{\pi n}/v_{F}$ where $v_F=10^6$m/s, and for BLG we use $m = 0.03 m_e$ \cite{li2016_blgmass}. Fig. 3a-d plots the density and temperature dependence of $\nu$ measured for both MLG and BLG. 
In MLG, $\nu$ follows a $T^{-1}$ dependence for nearly a decade of response, and shows no density dependence, within the measurement uncertainty  (Fig. 3a,c). The effective electron-electron scattering rate, which is given by $\tau_{ee}^{-1}=v_F^2/4\nu$, consequently has a $T$-linear dependence and, since $v_{F}$ is constant for MLG, is density-independent (Fig. 3e). In BLG, $\nu$ again follows $T^{-1}$ behaviour, but in this case varies linearly with $n$ (Fig. 3b,d). For the massive BLG band, $v_{F}=\hbar\sqrt{2\pi n}/m$ so that  $\nu\sim n/T$ translates again to an overall $T$-linear and density-independent scattering rate (Fig. 3f). 

The  observation that $\tau_{ee}^{-1}\sim T$ and is density independent for both massless electrons in MLG and massive electrons in BLG suggests that this scaling is independent of the band structure. 
Evidence for similar scaling can be seen in the data from previous graphene studies using different techniques,  but was not explicitly discussed \cite{Bandurin2016,Berdyugin2019,kumar2017superballistic}. According to Fermi liquid theory, the $e-e$ scattering rate is expected to scale as $\tau_{ee}^{-1}\sim T^2/E_F$ in two dimensions \cite{principi2016}. The discrepancy between the temperature and density dependences measured in both MLG and BLG from  conventional Fermi liquid theory is surprising and may urge careful reexamination of the standard Fermi liquid description.  
Recent theoretical work has identified that interacting 2D charged fluids may exhibit an exotic new regime between ballistic transport and hydrodynamic flow.  Named ``tomographic transport'', this new regime is characterized by a viscous type flow but where the lifetime of odd and even harmonic modes contribute differently to viscosity \cite{ledwith2019, Hofmann2022collectivemodes, Hofmann2023longlifetimes,Kryhin2023}
Within the tomographic flow regime, the effective scattering rate is predicted to go as  $\tau_{ee}^{-1}\sim T/n^{1/6}$ down to the lowest temperatures \cite{Kryhin2023}. The tomographic theory therefore aligns exactly with our observed $\tau_{ee}^{-1}\sim T$ behavior, and predicts a $n$ dependence much closer to the experimental observation (see SI for derivation and estimate of viscosity using tomographic transport theory). 

\head{Crossover betwen Fermi liquid and Dirac fluid}
So far we have examined charge densities sufficiently far from the charge neutrality point (CNP) to be considered a homogeneous Fermi liquid, and have considered a mixed flow regime involving both viscous and Ohmic components\cite{LevchenkoTransportPropertiesOf2020,shavit2019_corbino}.  At low densities close to the CNP ($E_F\ll k_B T$), graphene resembles instead a relativistic electron-hole plasma known as a Dirac fluid.  In this limit, transport is described by relativsitic hydrodynamics, governed by a combination of both $e$-$e$ and $e$-$h$ interactions. It was proposed that at the CNP,  graphene demonstrates a nearly ``perfect fluid'' where the ratio of shear viscosity, $\eta$,  to entropy density, $s$, approaches a universal quantum limit $\eta/s \ge \hbar/4\pi k_{B}$\cite{kovtun2005viscositybound,son2007prl,son2007arnpp,KARSCH2008,Sachdev2009_JPCM,muller2009perfectfluid}. 
Prior experimental studies identified evidence of hydrodynamic behaviour near the CNP\cite{Crossno2016,Ku2020NVhydro}, and with estimates of $\eta/s$ at room temperature reaching $\sim0.3\hbar/k_{B}$, close to the theoretical value $0.2\hbar/k_{B}$ predicted for MLG at 300~K\cite{muller2009perfectfluid,Ku2020NVhydro}. In the $e$-$h$ plasma regime, $\eta(T)$ is expected to show opposite behaviour to the Fermi liquid regime i.e. increasing, instead of decreasing, with temperature - providing an unambiguous and hall mark signature of the Dirac fluid. However, the temperature dependence of $\eta$ in the graphene electron-hole plasma regime has not previously been reported. Moreover, the crossover regime, $E_F\sim k_B T$,  where gate-induced carriers (Fermi liquid) and thermally excited carriers (Dirac fluid) coexist remains unexplored.

Fig. 4a shows $\eta$ versus $T$ measured at different carrier densities in MLG. The high density/low $T$ behaviour is well fit by  $\eta_{ee}^{MLG}\sim n^{3/2}/T$, consistent with the $\tau_{ee}^{-1}\sim T$ discussed above. By comparison, the low-$n$/high-$T$ behaviour collapses to a universal, density independent, curve where $\eta_{eh}^{MLG}\sim T^2$ in agreement with predictions for the charged Dirac fluid for MLG\cite{muller2009perfectfluid}. Assuming $E_F = k_B T$, we estimate $\nu\approx 0.3$ m$^2/$s for CNP at 300 K, consistent with result in recent magnetometry imaging experiment\cite{Ku2020NVhydro}. Analogous behaviour is observed in BLG (Fig. 4b) with $\eta(T)$ transitioning from $\eta_{ee}^{BLG}\sim n^{2}/T$ dependence at high density/low temperature, to a universal curve at low density/high temperature, $\eta_{eh}^{BLG}\sim T$, consistent with $\tau_{ee}^{-1}\sim T$, and  Dirac fluid theory for BLG\cite{yudhistira2023nonmonotonic}, respectively. 
In both MLG and BLG, the shear viscosity continuously interpolates between the Fermi liquid and Dirac fluid response as we vary $n$ or $T$. The full temperature range can be described by simply adding the contribution from gate-injected and thermally excited carriers, shown as solid curves in Fig. 4a,b (see SI for determination of the fitting coefficients).
\begin{equation}
 \eta=\eta_{eh} +\eta_{ee}   
\end{equation}.

This result suggest a two-fluid model where the gate-induced electrons and thermally excited electron-hole plasma coexist and contribute to hydrodynamic electrical transport in two parallel channels\cite{yudhistira2023nonmonotonic}. 
This allows us to construct a phase diagram based on the percentage contribution from electron-hole plasma in total hydrodynamic transport (Fig. 4c,d, plotted in Fermi energy units using the MLG and BLG band dispersion, respectively). 
 The data points label the crossover temperature, identified by the $\eta$ minimum in Fig. 4a,b.  The crossover boundary between the degenerate Fermi liquid and electron-hole plasma regimes follows approximately $E_{F}=2.3k_{B} T$ for MLG and $E_{F}=1.4k_{B} T$ for BLG.

Finally, from our measurement of $\eta(T)$, and using a calculated value for the entropy density from bandstsructure, $s(T)$\cite{muller2009perfectfluid}, we can determine the ratio $\eta/s$ as a function of temperature.  We find that for both BLG and MLG the ratio saturates in the low density high temperature limit, as expected (see SI). For MLG,  $\eta/s\approx 0.2\hbar/k_B$, consistent with theoretical predictions~\cite{muller2009perfectfluid} and previous experimental value estimated from profile imaging at 300~K\cite{Ku2020NVhydro}. For BLG the measured ratio is approximately four times as large~\cite{yudhistira2023nonmonotonic}.

\head{Summary}
\noindent
We demonstrate that the magnetoresistance in a Corbino geometry can be utilized to quantify the electron hydrodynamic viscosity, and therefore interparticle scattering rate in a 2D electron system.  Applying this technique to the study of monolayer and bilayer graphene we confirm the persistence of viscous effects over large range of density and temperature, and for the first time demonstrate the ability to quantify the Ohmic and viscous contributions in the crossover regime. 
 Our finding that temperature and density scaling of  of viscosity is at odds with Fermi liquid theory provides new  insight into the microscopic nature of interparticle scattering in a charged two-dimensional fluid.  The Corbino viscometer geometry demonstrated here can be universally applied to any 2d material system where the $e$-$e$ mean free path represents the dominant scattering length scale. A particularly promising application of hydrodynamic theory is understanding transport response in flat band systems with strong correlations.  Here strong $e$-$e$ interactions play a dominant role in determining ground state behaviour, but, reduced mobility due to the quenched kinetic energy makes it difficult within experimental parameters to eliminate Ohmic considerations and isolate the hydrodynamic components. The capability to disentangle the resistivity and viscosity parameters could therefore provide significant new opportunity to understanding these materials.

\section{Method}
\subsection{Device fabrication}
The heterostructure is assembled using standard van der Waals transfer technique. The device area is defined using e-beam lithography and shaped using consecutive SF$_6$/O$_2$ plasma etching. A small opening (width $\sim r_a/4$) is left out on the outer circumference for easy access to the inner electrode. Cr/Pd is deposited using e-beam evaporator to make one-dimensional contact\cite{Wang2013}. All MLG (BLG) devices shown in the this work are fabricated on the same heterostructure consisting of MLG (BLG) to ensure almost identical scattering length among devices. The carrier density $n$ at each gate voltage is determined from Shubnikov-de Haas oscillation. The hBN dielectric for both MLG and BLG devices are $\sim 50$ nm thick.

\subsection{Electrical transport measurements}
We perform electrical measurement using Stanford Research SR830 lock-in amplifier at a frequency of 77.77~Hz. Current flow through each device is less than $500$~nA to ensures a linear $I-V$ response\cite{Bandurin2016}. The higher frequency and larger current used here comparing to quantum transport measurement at much lower temperatures\cite{Zeng2019} is to gain a better signal to noise ratio. Our measurement configuration is consistent with that in previous experiments studying hydrodynamics in graphene \cite{Bandurin2016,Kumar2022LS,kumar2017superballistic,Ku2020NVhydro}.
Since multiples devices are sitting on the same substrate and are electrically isolated from each other, they are probed in parallel. An excitation voltage is applied to the inner electrodes of multiples devices simultaneously using one lock-in amplifier. Current drained out of the outer electrodes for each device is measured using separate lock-in amplifiers. 

\subsection{Magnetoresistance derived from the Stokes-Ohm equation}
The magnetotransport of electrons in the hydrodynamic-ohmic crossover regime is described by the Stokes-Ohm equation: $ne\boldsymbol{E}+\boldsymbol{j}\times\boldsymbol{B} = -\eta\Delta\dfrac{\boldsymbol{j}}{ne}+\rho_0\boldsymbol{j}ne$. In the coordinate of $(r,\theta)$, it writes:
\begin{align*}
    ne(E_r+\dfrac{j_\theta B}{ne})=-\eta\Delta\dfrac{j_r}{ne}+\rho_o j_r ne\\
    -ne\dfrac{j_r}{ne}B=-\eta\Delta\dfrac{j_\theta}{ne}+\rho_o j_\theta ne
\end{align*}
Where $E_r$ and $E_\theta$ are the radial and azimuthal electric field, $j_r$ and $j_\theta$ are the radial and azimuthal electrical current. Considering the rotational symmetry of the corbino disk, $E_r$, $E_\theta$, $j_r$ and $j_\theta$ depend on $r$ but not $\theta$. $\Delta\equiv\frac{\partial^2}{\partial r^2}+\frac{1}{r}\frac{\partial}{\partial r}-\frac{1}{r^2}$. In the corbino measurement setup, $E_\theta=0$. And the radial current is known from the charge conservation law: $j_r=\frac{I}{2\pi r}$. This gives us the following equation for the azimuthal current:
\begin{equation*}
    \frac{BI}{2\pi r}=\frac{\eta}{ne}(\frac{d^2 j_\theta}{dr^2}+\frac{1}{r}\frac{\partial j_\theta}{\partial r}-\frac{1}{r^2}j_\theta)+\rho_o nej_\theta
\end{equation*}
The general solution takes the form
\begin{equation*}
    j_\theta(r)=\frac{BI}{2\pi\rho_o ne}[AI_1(r/D_\nu)+BK_1(r/D_\nu)-\frac{1}{r}]
\end{equation*}
$I_1$ and $K_1$ are the modified Bessel functions of the second kind. Assuming the no-slip boundary condition, the normal velocity at the edge of corbino equals zero: $j_\theta(r_a)=j_\theta(r_b)=0$. One finds the coefficients:
\begin{align*}
    A=\frac{1}{C}(\frac{1}{r_a}K_1(\lambda)-\frac{1}{r_b}K_1(\lambda/\gamma))\\
    B=-\frac{1}{C}(\frac{1}{r_a}I_1(\lambda)-\frac{1}{r_b}I_1(\lambda/\gamma))\\
\end{align*}
where $C=I_1(\lambda/\gamma)K_1(\lambda)-I_1(\lambda)K_1(\lambda/\gamma)$, $r_b/D_\nu=\lambda$.
We then obtain the $B$-dependent radial electric field:
\begin{equation*}
    E_r(B) = \frac{B^2}{2\pi\rho (ne)^2}[\frac{1}{r}-AI_1(r/D_\nu)-BK_1(r/D_\nu)]
\end{equation*}
The potential drop between the inner and outer electrode is evaluated by integrating $E_r$.
$\Delta U= \int_{r_a}^{r_b} E_r dr$.
Finally, the magnetoresistance is obtained by dividing $\Delta U$ with I (Eq. 1 in the main text):
\begin{equation*}
    R(B) = \frac{B^2 \ln\gamma}{2\pi\rho_o(ne)^2}[1-f(\gamma,\lambda)]
\end{equation*}
where $f(\gamma,\lambda)=$
\begin{multline*}
\frac{1}{\ln\gamma}\Bigl\{\frac{[I_0(\lambda)-I_0(\lambda/\gamma)][(\gamma /\lambda)K_1(\lambda)-(1/\lambda)K_1(\lambda/\gamma)]}{I_1(\lambda/\gamma)K_1(\lambda)-I_1(\lambda)K_1(\lambda/\gamma)}\\
+\frac{[K_0(\lambda)-K_0(\lambda/\gamma)][(\gamma /\lambda)I_1(\lambda)-(1/\lambda)I_1(\lambda/\gamma)]}{I_1(\lambda/\gamma)K_1(\lambda)-I_1(\lambda)K_1(\lambda/\gamma)}\Bigr\}
\end{multline*}
$I_0$ and $K_0$ are the modified Bessel functions of the first kind.

\section{acknowledgments}
We thank Zhiyuan Sun, Alex Levchenko, Andrey Shytov, Shahal Ilani, Shaffique Adam for helpful discussion. YZ thanks Maelle Kapfer for assistance in data analysis. 

This research is primarily supported by the Columbia
University Materials Science and Engineering Research
Center (MRSEC), through NSF grants DMR-2011738. K.W. and T.T. acknowledge support from the JSPS KAKENHI (Grant Numbers 21H05233 and 23H02052) and World Premier International Research Center Initiative (WPI), MEXT, Japan. The theory part of this work at MIT was supported by the Science and Technology Center for Integrated Quantum Materials, National Science Foundation grant No.\,DMR1231319; US-Israel Binational Science Foundation Grant No.\,2018033.

\section{Data availability}
The data that support the plots within this paper and other findings of this study are available from the corresponding author upon reasonable request.

\section{Author contribution}
Y.Z. and C.R.D. conceived the experiment. Y.Z.and O.M.G. fabricated the devices. Y.Z. performed the measurement and data analysis. H.G and L.L provides theory model. Y.Z., L.L. and C.R.D. wrote the manuscript with input from all authors.

\section*{Competing financial interests}
The authors declare no competing financial interests.

\bibliography{YZ_ref}%

\begin{thebibliography}{55}%
\makeatletter
\providecommand \@ifxundefined [1]{%
 \@ifx{#1\undefined}
}%
\providecommand \@ifnum [1]{%
 \ifnum #1\expandafter \@firstoftwo
 \else \expandafter \@secondoftwo
 \fi
}%
\providecommand \@ifx [1]{%
 \ifx #1\expandafter \@firstoftwo
 \else \expandafter \@secondoftwo
 \fi
}%
\providecommand \natexlab [1]{#1}%
\providecommand \enquote  [1]{``#1''}%
\providecommand \bibnamefont  [1]{#1}%
\providecommand \bibfnamefont [1]{#1}%
\providecommand \citenamefont [1]{#1}%
\providecommand \href@noop [0]{\@secondoftwo}%
\providecommand \href [0]{\begingroup \@sanitize@url \@href}%
\providecommand \@href[1]{\@@startlink{#1}\@@href}%
\providecommand \@@href[1]{\endgroup#1\@@endlink}%
\providecommand \@sanitize@url [0]{\catcode `\\12\catcode `\$12\catcode `\&12\catcode `\#12\catcode `\^12\catcode `\_12\catcode `\%12\relax}%
\providecommand \@@startlink[1]{}%
\providecommand \@@endlink[0]{}%
\providecommand \url  [0]{\begingroup\@sanitize@url \@url }%
\providecommand \@url [1]{\endgroup\@href {#1}{\urlprefix }}%
\providecommand \urlprefix  [0]{URL }%
\providecommand \Eprint [0]{\href }%
\providecommand \doibase [0]{http://dx.doi.org/}%
\providecommand \selectlanguage [0]{\@gobble}%
\providecommand \bibinfo  [0]{\@secondoftwo}%
\providecommand \bibfield  [0]{\@secondoftwo}%
\providecommand \translation [1]{[#1]}%
\providecommand \BibitemOpen [0]{}%
\providecommand \bibitemStop [0]{}%
\providecommand \bibitemNoStop [0]{.\EOS\space}%
\providecommand \EOS [0]{\spacefactor3000\relax}%
\providecommand \BibitemShut  [1]{\csname bibitem#1\endcsname}%
\let\auto@bib@innerbib\@empty
\bibitem [{\citenamefont {Alexandradinata}\ \emph {et~al.}(2022)\citenamefont {Alexandradinata}, \citenamefont {Armitage}, \citenamefont {Baydin}, \citenamefont {Bi}, \citenamefont {Cao}, \citenamefont {Changlani}, \citenamefont {Chertkov}, \citenamefont {da~Silva~Neto}, \citenamefont {Delacretaz}, \citenamefont {Baggari}, \citenamefont {Ferguson}, \citenamefont {Gannon}, \citenamefont {Ghorashi}, \citenamefont {Goodge}, \citenamefont {Goulko}, \citenamefont {Grissonnanche}, \citenamefont {Hallas}, \citenamefont {Hayes}, \citenamefont {He}, \citenamefont {Huang}, \citenamefont {Kogar}, \citenamefont {Kumah}, \citenamefont {Lee}, \citenamefont {Legros}, \citenamefont {Mahmood}, \citenamefont {Maximenko}, \citenamefont {Pellatz}, \citenamefont {Polshyn}, \citenamefont {Sarkar}, \citenamefont {Scheie}, \citenamefont {Seyler}, \citenamefont {Shi}, \citenamefont {Skinner}, \citenamefont {Steinke}, \citenamefont {Thirunavukkuarasu}, \citenamefont {Trevisan}, \citenamefont {Vogl}, \citenamefont {Volkov},
  \citenamefont {Wang}, \citenamefont {Wang}, \citenamefont {Wei}, \citenamefont {Wei}, \citenamefont {Yang}, \citenamefont {Zhang}, \citenamefont {Zhang}, \citenamefont {Zhao},\ and\ \citenamefont {Zong}}]{alexandradinata2022futurecorrelatedelectronproblem}%
  \BibitemOpen
  \bibfield  {author} {\bibinfo {author} {\bibfnamefont {A.}~\bibnamefont {Alexandradinata}}, \bibinfo {author} {\bibfnamefont {N.~P.}\ \bibnamefont {Armitage}}, \bibinfo {author} {\bibfnamefont {A.}~\bibnamefont {Baydin}}, \bibinfo {author} {\bibfnamefont {W.}~\bibnamefont {Bi}}, \bibinfo {author} {\bibfnamefont {Y.}~\bibnamefont {Cao}}, \bibinfo {author} {\bibfnamefont {H.~J.}\ \bibnamefont {Changlani}}, \bibinfo {author} {\bibfnamefont {E.}~\bibnamefont {Chertkov}}, \bibinfo {author} {\bibfnamefont {E.~H.}\ \bibnamefont {da~Silva~Neto}}, \bibinfo {author} {\bibfnamefont {L.}~\bibnamefont {Delacretaz}}, \bibinfo {author} {\bibfnamefont {I.~E.}\ \bibnamefont {Baggari}}, \bibinfo {author} {\bibfnamefont {G.~M.}\ \bibnamefont {Ferguson}}, \bibinfo {author} {\bibfnamefont {W.~J.}\ \bibnamefont {Gannon}}, \bibinfo {author} {\bibfnamefont {S.~A.~A.}\ \bibnamefont {Ghorashi}}, \bibinfo {author} {\bibfnamefont {B.~H.}\ \bibnamefont {Goodge}}, \bibinfo {author} {\bibfnamefont {O.}~\bibnamefont {Goulko}}, \bibinfo
  {author} {\bibfnamefont {G.}~\bibnamefont {Grissonnanche}}, \bibinfo {author} {\bibfnamefont {A.}~\bibnamefont {Hallas}}, \bibinfo {author} {\bibfnamefont {I.~M.}\ \bibnamefont {Hayes}}, \bibinfo {author} {\bibfnamefont {Y.}~\bibnamefont {He}}, \bibinfo {author} {\bibfnamefont {E.~W.}\ \bibnamefont {Huang}}, \bibinfo {author} {\bibfnamefont {A.}~\bibnamefont {Kogar}}, \bibinfo {author} {\bibfnamefont {D.}~\bibnamefont {Kumah}}, \bibinfo {author} {\bibfnamefont {J.~Y.}\ \bibnamefont {Lee}}, \bibinfo {author} {\bibfnamefont {A.}~\bibnamefont {Legros}}, \bibinfo {author} {\bibfnamefont {F.}~\bibnamefont {Mahmood}}, \bibinfo {author} {\bibfnamefont {Y.}~\bibnamefont {Maximenko}}, \bibinfo {author} {\bibfnamefont {N.}~\bibnamefont {Pellatz}}, \bibinfo {author} {\bibfnamefont {H.}~\bibnamefont {Polshyn}}, \bibinfo {author} {\bibfnamefont {T.}~\bibnamefont {Sarkar}}, \bibinfo {author} {\bibfnamefont {A.}~\bibnamefont {Scheie}}, \bibinfo {author} {\bibfnamefont {K.~L.}\ \bibnamefont {Seyler}}, \bibinfo {author}
  {\bibfnamefont {Z.}~\bibnamefont {Shi}}, \bibinfo {author} {\bibfnamefont {B.}~\bibnamefont {Skinner}}, \bibinfo {author} {\bibfnamefont {L.}~\bibnamefont {Steinke}}, \bibinfo {author} {\bibfnamefont {K.}~\bibnamefont {Thirunavukkuarasu}}, \bibinfo {author} {\bibfnamefont {T.~V.}\ \bibnamefont {Trevisan}}, \bibinfo {author} {\bibfnamefont {M.}~\bibnamefont {Vogl}}, \bibinfo {author} {\bibfnamefont {P.~A.}\ \bibnamefont {Volkov}}, \bibinfo {author} {\bibfnamefont {Y.}~\bibnamefont {Wang}}, \bibinfo {author} {\bibfnamefont {Y.}~\bibnamefont {Wang}}, \bibinfo {author} {\bibfnamefont {D.}~\bibnamefont {Wei}}, \bibinfo {author} {\bibfnamefont {K.}~\bibnamefont {Wei}}, \bibinfo {author} {\bibfnamefont {S.}~\bibnamefont {Yang}}, \bibinfo {author} {\bibfnamefont {X.}~\bibnamefont {Zhang}}, \bibinfo {author} {\bibfnamefont {Y.-H.}\ \bibnamefont {Zhang}}, \bibinfo {author} {\bibfnamefont {L.}~\bibnamefont {Zhao}}, \ and\ \bibinfo {author} {\bibfnamefont {A.}~\bibnamefont {Zong}},\ }\href
  {https://arxiv.org/abs/2010.00584} {\  (\bibinfo {year} {2022})},\ \Eprint {http://arxiv.org/abs/2010.00584} {arXiv:2010.00584 [cond-mat.str-el]} \BibitemShut {NoStop}%
\bibitem [{\citenamefont {de~Jong}\ and\ \citenamefont {Molenkamp}(1995)}]{Molenkamp1995prb}%
  \BibitemOpen
  \bibfield  {author} {\bibinfo {author} {\bibfnamefont {M.~J.~M.}\ \bibnamefont {de~Jong}}\ and\ \bibinfo {author} {\bibfnamefont {L.~W.}\ \bibnamefont {Molenkamp}},\ }\href {\doibase 10.1103/PhysRevB.51.13389} {\bibfield  {journal} {\bibinfo  {journal} {Phys. Rev. B}\ }\textbf {\bibinfo {volume} {51}},\ \bibinfo {pages} {13389} (\bibinfo {year} {1995})}\BibitemShut {NoStop}%
\bibitem [{\citenamefont {Damle}\ and\ \citenamefont {Sachdev}(1997)}]{damle1997}%
  \BibitemOpen
  \bibfield  {author} {\bibinfo {author} {\bibfnamefont {K.}~\bibnamefont {Damle}}\ and\ \bibinfo {author} {\bibfnamefont {S.}~\bibnamefont {Sachdev}},\ }\href {\doibase 10.1103/PhysRevB.56.8714} {\bibfield  {journal} {\bibinfo  {journal} {Phys. Rev. B}\ }\textbf {\bibinfo {volume} {56}},\ \bibinfo {pages} {8714} (\bibinfo {year} {1997})}\BibitemShut {NoStop}%
\bibitem [{\citenamefont {Davison}\ \emph {et~al.}(2014)\citenamefont {Davison}, \citenamefont {Schalm},\ and\ \citenamefont {Zaanen}}]{Davison2014}%
  \BibitemOpen
  \bibfield  {author} {\bibinfo {author} {\bibfnamefont {R.~A.}\ \bibnamefont {Davison}}, \bibinfo {author} {\bibfnamefont {K.}~\bibnamefont {Schalm}}, \ and\ \bibinfo {author} {\bibfnamefont {J.}~\bibnamefont {Zaanen}},\ }\href {\doibase 10.1103/PhysRevB.89.245116} {\bibfield  {journal} {\bibinfo  {journal} {Phys. Rev. B}\ }\textbf {\bibinfo {volume} {89}},\ \bibinfo {pages} {245116} (\bibinfo {year} {2014})}\BibitemShut {NoStop}%
\bibitem [{\citenamefont {Forcella}\ \emph {et~al.}(2014)\citenamefont {Forcella}, \citenamefont {Zaanen}, \citenamefont {Valentinis},\ and\ \citenamefont {van~der Marel}}]{forcella2014}%
  \BibitemOpen
  \bibfield  {author} {\bibinfo {author} {\bibfnamefont {D.}~\bibnamefont {Forcella}}, \bibinfo {author} {\bibfnamefont {J.}~\bibnamefont {Zaanen}}, \bibinfo {author} {\bibfnamefont {D.}~\bibnamefont {Valentinis}}, \ and\ \bibinfo {author} {\bibfnamefont {D.}~\bibnamefont {van~der Marel}},\ }\href {\doibase 10.1103/PhysRevB.90.035143} {\bibfield  {journal} {\bibinfo  {journal} {Phys. Rev. B}\ }\textbf {\bibinfo {volume} {90}},\ \bibinfo {pages} {035143} (\bibinfo {year} {2014})}\BibitemShut {NoStop}%
\bibitem [{\citenamefont {Levitov}\ and\ \citenamefont {Falkovich}(2016)}]{Levitov2016}%
  \BibitemOpen
  \bibfield  {author} {\bibinfo {author} {\bibfnamefont {L.}~\bibnamefont {Levitov}}\ and\ \bibinfo {author} {\bibfnamefont {G.}~\bibnamefont {Falkovich}},\ }\href {\doibase 10.1038/nphys3667} {\bibfield  {journal} {\bibinfo  {journal} {Nature Physics}\ }\textbf {\bibinfo {volume} {12}},\ \bibinfo {pages} {672} (\bibinfo {year} {2016})}\BibitemShut {NoStop}%
\bibitem [{\citenamefont {Gurzhi}(1963)}]{gurzhi1962nonmonotonic}%
  \BibitemOpen
  \bibfield  {author} {\bibinfo {author} {\bibfnamefont {R.}~\bibnamefont {Gurzhi}},\ }\href@noop {} {\bibfield  {journal} {\bibinfo  {journal} {Sov. Phys. JETP}\ }\textbf {\bibinfo {volume} {44}},\ \bibinfo {pages} {771} (\bibinfo {year} {1963})}\BibitemShut {NoStop}%
\bibitem [{\citenamefont {Gurzhi}(1968)}]{Gurzhi_1968}%
  \BibitemOpen
  \bibfield  {author} {\bibinfo {author} {\bibfnamefont {R.~N.}\ \bibnamefont {Gurzhi}},\ }\href {\doibase 10.1070/pu1968v011n02abeh003815} {\bibfield  {journal} {\bibinfo  {journal} {Soviet Physics Uspekhi}\ }\textbf {\bibinfo {volume} {11}},\ \bibinfo {pages} {255} (\bibinfo {year} {1968})}\BibitemShut {NoStop}%
\bibitem [{\citenamefont {M\"uller}\ \emph {et~al.}(2009)\citenamefont {M\"uller}, \citenamefont {Schmalian},\ and\ \citenamefont {Fritz}}]{muller2009perfectfluid}%
  \BibitemOpen
  \bibfield  {author} {\bibinfo {author} {\bibfnamefont {M.}~\bibnamefont {M\"uller}}, \bibinfo {author} {\bibfnamefont {J.}~\bibnamefont {Schmalian}}, \ and\ \bibinfo {author} {\bibfnamefont {L.}~\bibnamefont {Fritz}},\ }\href {\doibase 10.1103/PhysRevLett.103.025301} {\bibfield  {journal} {\bibinfo  {journal} {Phys. Rev. Lett.}\ }\textbf {\bibinfo {volume} {103}},\ \bibinfo {pages} {025301} (\bibinfo {year} {2009})}\BibitemShut {NoStop}%
\bibitem [{\citenamefont {Ledwith}\ \emph {et~al.}(2019{\natexlab{a}})\citenamefont {Ledwith}, \citenamefont {Guo}, \citenamefont {Shytov},\ and\ \citenamefont {Levitov}}]{ledwith2019}%
  \BibitemOpen
  \bibfield  {author} {\bibinfo {author} {\bibfnamefont {P.}~\bibnamefont {Ledwith}}, \bibinfo {author} {\bibfnamefont {H.}~\bibnamefont {Guo}}, \bibinfo {author} {\bibfnamefont {A.}~\bibnamefont {Shytov}}, \ and\ \bibinfo {author} {\bibfnamefont {L.}~\bibnamefont {Levitov}},\ }\href {\doibase 10.1103/PhysRevLett.123.116601} {\bibfield  {journal} {\bibinfo  {journal} {Phys. Rev. Lett.}\ }\textbf {\bibinfo {volume} {123}},\ \bibinfo {pages} {116601} (\bibinfo {year} {2019}{\natexlab{a}})}\BibitemShut {NoStop}%
\bibitem [{\citenamefont {Ledwith}\ \emph {et~al.}(2019{\natexlab{b}})\citenamefont {Ledwith}, \citenamefont {Guo},\ and\ \citenamefont {Levitov}}]{ledwith2019hierarchy}%
  \BibitemOpen
  \bibfield  {author} {\bibinfo {author} {\bibfnamefont {P.~J.}\ \bibnamefont {Ledwith}}, \bibinfo {author} {\bibfnamefont {H.}~\bibnamefont {Guo}}, \ and\ \bibinfo {author} {\bibfnamefont {L.}~\bibnamefont {Levitov}},\ }\href {\doibase https://doi.org/10.1016/j.aop.2019.167913} {\bibfield  {journal} {\bibinfo  {journal} {Annals of Physics}\ }\textbf {\bibinfo {volume} {411}},\ \bibinfo {pages} {167913} (\bibinfo {year} {2019}{\natexlab{b}})}\BibitemShut {NoStop}%
\bibitem [{\citenamefont {Hofmann}\ and\ \citenamefont {Gran}(2023)}]{Hofmann2023longlifetimes}%
  \BibitemOpen
  \bibfield  {author} {\bibinfo {author} {\bibfnamefont {J.}~\bibnamefont {Hofmann}}\ and\ \bibinfo {author} {\bibfnamefont {U.}~\bibnamefont {Gran}},\ }\href {\doibase 10.1103/PhysRevB.108.L121401} {\bibfield  {journal} {\bibinfo  {journal} {Phys. Rev. B}\ }\textbf {\bibinfo {volume} {108}},\ \bibinfo {pages} {L121401} (\bibinfo {year} {2023})}\BibitemShut {NoStop}%
\bibitem [{\citenamefont {Hofmann}\ and\ \citenamefont {Das~Sarma}(2022)}]{Hofmann2022collectivemodes}%
  \BibitemOpen
  \bibfield  {author} {\bibinfo {author} {\bibfnamefont {J.}~\bibnamefont {Hofmann}}\ and\ \bibinfo {author} {\bibfnamefont {S.}~\bibnamefont {Das~Sarma}},\ }\href {\doibase 10.1103/PhysRevB.106.205412} {\bibfield  {journal} {\bibinfo  {journal} {Phys. Rev. B}\ }\textbf {\bibinfo {volume} {106}},\ \bibinfo {pages} {205412} (\bibinfo {year} {2022})}\BibitemShut {NoStop}%
\bibitem [{\citenamefont {Kryhin}\ \emph {et~al.}(2023)\citenamefont {Kryhin}, \citenamefont {Hong},\ and\ \citenamefont {Levitov}}]{Kryhin2023}%
  \BibitemOpen
  \bibfield  {author} {\bibinfo {author} {\bibfnamefont {S.}~\bibnamefont {Kryhin}}, \bibinfo {author} {\bibfnamefont {Q.}~\bibnamefont {Hong}}, \ and\ \bibinfo {author} {\bibfnamefont {L.}~\bibnamefont {Levitov}},\ }\href@noop {} {\  (\bibinfo {year} {2023})},\ \Eprint {http://arxiv.org/abs/2310.08556} {arxiv:2310.08556 [cond-mat]} \BibitemShut {NoStop}%
\bibitem [{\citenamefont {Nilsson}\ \emph {et~al.}(2024)\citenamefont {Nilsson}, \citenamefont {Gran},\ and\ \citenamefont {Hofmann}}]{nilsson2024nonequilibriumrelaxationoddeveneffect}%
  \BibitemOpen
  \bibfield  {author} {\bibinfo {author} {\bibfnamefont {E.}~\bibnamefont {Nilsson}}, \bibinfo {author} {\bibfnamefont {U.}~\bibnamefont {Gran}}, \ and\ \bibinfo {author} {\bibfnamefont {J.}~\bibnamefont {Hofmann}},\ }\href {https://arxiv.org/abs/2405.03635} {\  (\bibinfo {year} {2024})},\ \Eprint {http://arxiv.org/abs/2405.03635} {arXiv:2405.03635 [cond-mat.mes-hall]} \BibitemShut {NoStop}%
\bibitem [{\citenamefont {Hong}\ \emph {et~al.}(2020)\citenamefont {Hong}, \citenamefont {Davydova}, \citenamefont {Ledwith},\ and\ \citenamefont {Levitov}}]{hong2020superscreeningretroreflectedholebackflow}%
  \BibitemOpen
  \bibfield  {author} {\bibinfo {author} {\bibfnamefont {Q.}~\bibnamefont {Hong}}, \bibinfo {author} {\bibfnamefont {M.}~\bibnamefont {Davydova}}, \bibinfo {author} {\bibfnamefont {P.~J.}\ \bibnamefont {Ledwith}}, \ and\ \bibinfo {author} {\bibfnamefont {L.}~\bibnamefont {Levitov}},\ }\href {https://arxiv.org/abs/2012.03840} {\  (\bibinfo {year} {2020})},\ \Eprint {http://arxiv.org/abs/2012.03840} {arXiv:2012.03840 [cond-mat.mes-hall]} \BibitemShut {NoStop}%
\bibitem [{\citenamefont {Eisenstein}\ \emph {et~al.}(2007)\citenamefont {Eisenstein}, \citenamefont {Syphers}, \citenamefont {Pfeiffer},\ and\ \citenamefont {West}}]{Eisenstein2007quantumlifetime}%
  \BibitemOpen
  \bibfield  {author} {\bibinfo {author} {\bibfnamefont {J.}~\bibnamefont {Eisenstein}}, \bibinfo {author} {\bibfnamefont {D.}~\bibnamefont {Syphers}}, \bibinfo {author} {\bibfnamefont {L.}~\bibnamefont {Pfeiffer}}, \ and\ \bibinfo {author} {\bibfnamefont {K.}~\bibnamefont {West}},\ }\href {\doibase https://doi.org/10.1016/j.ssc.2007.06.010} {\bibfield  {journal} {\bibinfo  {journal} {Solid State Communications}\ }\textbf {\bibinfo {volume} {143}},\ \bibinfo {pages} {365} (\bibinfo {year} {2007})}\BibitemShut {NoStop}%
\bibitem [{\citenamefont {Murphy}\ \emph {et~al.}(1995)\citenamefont {Murphy}, \citenamefont {Eisenstein}, \citenamefont {Pfeiffer},\ and\ \citenamefont {West}}]{Murphy1995lifetimetunneling}%
  \BibitemOpen
  \bibfield  {author} {\bibinfo {author} {\bibfnamefont {S.~Q.}\ \bibnamefont {Murphy}}, \bibinfo {author} {\bibfnamefont {J.~P.}\ \bibnamefont {Eisenstein}}, \bibinfo {author} {\bibfnamefont {L.~N.}\ \bibnamefont {Pfeiffer}}, \ and\ \bibinfo {author} {\bibfnamefont {K.~W.}\ \bibnamefont {West}},\ }\href {\doibase 10.1103/PhysRevB.52.14825} {\bibfield  {journal} {\bibinfo  {journal} {Phys. Rev. B}\ }\textbf {\bibinfo {volume} {52}},\ \bibinfo {pages} {14825} (\bibinfo {year} {1995})}\BibitemShut {NoStop}%
\bibitem [{\citenamefont {Lucas}\ and\ \citenamefont {Fong}(2018)}]{lucas2018reviewhydrodynamics}%
  \BibitemOpen
  \bibfield  {author} {\bibinfo {author} {\bibfnamefont {A.}~\bibnamefont {Lucas}}\ and\ \bibinfo {author} {\bibfnamefont {K.~C.}\ \bibnamefont {Fong}},\ }\href {\doibase 10.1088/1361-648X/aaa274} {\bibfield  {journal} {\bibinfo  {journal} {Journal of Physics: Condensed Matter}\ }\textbf {\bibinfo {volume} {30}},\ \bibinfo {pages} {053001} (\bibinfo {year} {2018})}\BibitemShut {NoStop}%
\bibitem [{\citenamefont {Levchenko}\ and\ \citenamefont {Schmalian}(2020)}]{LevchenkoTransportPropertiesOf2020}%
  \BibitemOpen
  \bibfield  {author} {\bibinfo {author} {\bibfnamefont {A.}~\bibnamefont {Levchenko}}\ and\ \bibinfo {author} {\bibfnamefont {J.}~\bibnamefont {Schmalian}},\ }\href {\doibase 10.1016/j.aop.2020.168218} {\bibfield  {journal} {\bibinfo  {journal} {Annals of Physics}\ }\textbf {\bibinfo {volume} {419}} (\bibinfo {year} {2020}),\ 10.1016/j.aop.2020.168218}\BibitemShut {NoStop}%
\bibitem [{\citenamefont {Molenkamp}\ and\ \citenamefont {de~Jong}(1994)}]{Molenkamp1994prb}%
  \BibitemOpen
  \bibfield  {author} {\bibinfo {author} {\bibfnamefont {L.~W.}\ \bibnamefont {Molenkamp}}\ and\ \bibinfo {author} {\bibfnamefont {M.~J.~M.}\ \bibnamefont {de~Jong}},\ }\href {\doibase 10.1103/PhysRevB.49.5038} {\bibfield  {journal} {\bibinfo  {journal} {Phys. Rev. B}\ }\textbf {\bibinfo {volume} {49}},\ \bibinfo {pages} {5038} (\bibinfo {year} {1994})}\BibitemShut {NoStop}%
\bibitem [{\citenamefont {Moll}\ \emph {et~al.}(2016)\citenamefont {Moll}, \citenamefont {Kushwaha}, \citenamefont {Nandi}, \citenamefont {Schmidt},\ and\ \citenamefont {Mackenzie}}]{Moll2016}%
  \BibitemOpen
  \bibfield  {author} {\bibinfo {author} {\bibfnamefont {P.~J.~W.}\ \bibnamefont {Moll}}, \bibinfo {author} {\bibfnamefont {P.}~\bibnamefont {Kushwaha}}, \bibinfo {author} {\bibfnamefont {N.}~\bibnamefont {Nandi}}, \bibinfo {author} {\bibfnamefont {B.}~\bibnamefont {Schmidt}}, \ and\ \bibinfo {author} {\bibfnamefont {A.~P.}\ \bibnamefont {Mackenzie}},\ }\href {\doibase 10.1126/science.aac8385} {\bibfield  {journal} {\bibinfo  {journal} {Science}\ }\textbf {\bibinfo {volume} {351}},\ \bibinfo {pages} {1061} (\bibinfo {year} {2016})}\BibitemShut {NoStop}%
\bibitem [{\citenamefont {Krishna~Kumar}\ \emph {et~al.}(2017)\citenamefont {Krishna~Kumar}, \citenamefont {Bandurin}, \citenamefont {Pellegrino}, \citenamefont {Cao}, \citenamefont {Principi}, \citenamefont {Guo}, \citenamefont {Auton}, \citenamefont {Ben~Shalom}, \citenamefont {Ponomarenko}, \citenamefont {Falkovich}, \citenamefont {Watanabe}, \citenamefont {Taniguchi}, \citenamefont {Grigorieva}, \citenamefont {Levitov}, \citenamefont {Polini},\ and\ \citenamefont {Geim}}]{kumar2017superballistic}%
  \BibitemOpen
  \bibfield  {author} {\bibinfo {author} {\bibfnamefont {R.}~\bibnamefont {Krishna~Kumar}}, \bibinfo {author} {\bibfnamefont {D.~A.}\ \bibnamefont {Bandurin}}, \bibinfo {author} {\bibfnamefont {F.~M.~D.}\ \bibnamefont {Pellegrino}}, \bibinfo {author} {\bibfnamefont {Y.}~\bibnamefont {Cao}}, \bibinfo {author} {\bibfnamefont {A.}~\bibnamefont {Principi}}, \bibinfo {author} {\bibfnamefont {H.}~\bibnamefont {Guo}}, \bibinfo {author} {\bibfnamefont {G.}~\bibnamefont {Auton}}, \bibinfo {author} {\bibfnamefont {M.}~\bibnamefont {Ben~Shalom}}, \bibinfo {author} {\bibfnamefont {L.~A.}\ \bibnamefont {Ponomarenko}}, \bibinfo {author} {\bibfnamefont {G.}~\bibnamefont {Falkovich}}, \bibinfo {author} {\bibfnamefont {K.}~\bibnamefont {Watanabe}}, \bibinfo {author} {\bibfnamefont {T.}~\bibnamefont {Taniguchi}}, \bibinfo {author} {\bibfnamefont {I.}~\bibnamefont {Grigorieva}}, \bibinfo {author} {\bibfnamefont {L.~S.}\ \bibnamefont {Levitov}}, \bibinfo {author} {\bibfnamefont {M.}~\bibnamefont {Polini}}, \ and\ \bibinfo
  {author} {\bibfnamefont {A.}~\bibnamefont {Geim}},\ }\href {\doibase 10.1038/nphys4240} {\bibfield  {journal} {\bibinfo  {journal} {Nature Physics}\ }\textbf {\bibinfo {volume} {13}},\ \bibinfo {pages} {1182} (\bibinfo {year} {2017})}\BibitemShut {NoStop}%
\bibitem [{\citenamefont {Ginzburg}\ \emph {et~al.}(2023)\citenamefont {Ginzburg}, \citenamefont {Wu}, \citenamefont {R\"o\"osli}, \citenamefont {Gomez}, \citenamefont {Garreis}, \citenamefont {Tong}, \citenamefont {Star\'a}, \citenamefont {Gold}, \citenamefont {Nazaryan}, \citenamefont {Kryhin}, \citenamefont {Overweg}, \citenamefont {Reichl}, \citenamefont {Berl}, \citenamefont {Taniguchi}, \citenamefont {Watanabe}, \citenamefont {Wegscheider}, \citenamefont {Ihn},\ and\ \citenamefont {Ensslin}}]{Ginzburg2023}%
  \BibitemOpen
  \bibfield  {author} {\bibinfo {author} {\bibfnamefont {L.~V.}\ \bibnamefont {Ginzburg}}, \bibinfo {author} {\bibfnamefont {Y.}~\bibnamefont {Wu}}, \bibinfo {author} {\bibfnamefont {M.~P.}\ \bibnamefont {R\"o\"osli}}, \bibinfo {author} {\bibfnamefont {P.~R.}\ \bibnamefont {Gomez}}, \bibinfo {author} {\bibfnamefont {R.}~\bibnamefont {Garreis}}, \bibinfo {author} {\bibfnamefont {C.}~\bibnamefont {Tong}}, \bibinfo {author} {\bibfnamefont {V.}~\bibnamefont {Star\'a}}, \bibinfo {author} {\bibfnamefont {C.}~\bibnamefont {Gold}}, \bibinfo {author} {\bibfnamefont {K.}~\bibnamefont {Nazaryan}}, \bibinfo {author} {\bibfnamefont {S.}~\bibnamefont {Kryhin}}, \bibinfo {author} {\bibfnamefont {H.}~\bibnamefont {Overweg}}, \bibinfo {author} {\bibfnamefont {C.}~\bibnamefont {Reichl}}, \bibinfo {author} {\bibfnamefont {M.}~\bibnamefont {Berl}}, \bibinfo {author} {\bibfnamefont {T.}~\bibnamefont {Taniguchi}}, \bibinfo {author} {\bibfnamefont {K.}~\bibnamefont {Watanabe}}, \bibinfo {author} {\bibfnamefont {W.}~\bibnamefont
  {Wegscheider}}, \bibinfo {author} {\bibfnamefont {T.}~\bibnamefont {Ihn}}, \ and\ \bibinfo {author} {\bibfnamefont {K.}~\bibnamefont {Ensslin}},\ }\href {\doibase 10.1103/PhysRevResearch.5.043088} {\bibfield  {journal} {\bibinfo  {journal} {Phys. Rev. Res.}\ }\textbf {\bibinfo {volume} {5}},\ \bibinfo {pages} {043088} (\bibinfo {year} {2023})}\BibitemShut {NoStop}%
\bibitem [{\citenamefont {Bandurin}\ \emph {et~al.}(2016)\citenamefont {Bandurin}, \citenamefont {Torre}, \citenamefont {Kumar}, \citenamefont {Ben~Shalom}, \citenamefont {Tomadin}, \citenamefont {Principi}, \citenamefont {Auton}, \citenamefont {Khestanova}, \citenamefont {Novoselov}, \citenamefont {Grigorieva}, \citenamefont {Ponomarenko}, \citenamefont {Geim},\ and\ \citenamefont {Polini}}]{Bandurin2016}%
  \BibitemOpen
  \bibfield  {author} {\bibinfo {author} {\bibfnamefont {D.~A.}\ \bibnamefont {Bandurin}}, \bibinfo {author} {\bibfnamefont {I.}~\bibnamefont {Torre}}, \bibinfo {author} {\bibfnamefont {R.~K.}\ \bibnamefont {Kumar}}, \bibinfo {author} {\bibfnamefont {M.}~\bibnamefont {Ben~Shalom}}, \bibinfo {author} {\bibfnamefont {A.}~\bibnamefont {Tomadin}}, \bibinfo {author} {\bibfnamefont {A.}~\bibnamefont {Principi}}, \bibinfo {author} {\bibfnamefont {G.~H.}\ \bibnamefont {Auton}}, \bibinfo {author} {\bibfnamefont {E.}~\bibnamefont {Khestanova}}, \bibinfo {author} {\bibfnamefont {K.~S.}\ \bibnamefont {Novoselov}}, \bibinfo {author} {\bibfnamefont {I.~V.}\ \bibnamefont {Grigorieva}}, \bibinfo {author} {\bibfnamefont {L.~A.}\ \bibnamefont {Ponomarenko}}, \bibinfo {author} {\bibfnamefont {A.~K.}\ \bibnamefont {Geim}}, \ and\ \bibinfo {author} {\bibfnamefont {M.}~\bibnamefont {Polini}},\ }\href {\doibase 10.1126/science.aad0201} {\bibfield  {journal} {\bibinfo  {journal} {Science}\ }\textbf {\bibinfo {volume} {351}},\
  \bibinfo {pages} {1055} (\bibinfo {year} {2016})}\BibitemShut {NoStop}%
\bibitem [{\citenamefont {Berdyugin}\ \emph {et~al.}(2019)\citenamefont {Berdyugin}, \citenamefont {Xu}, \citenamefont {Pellegrino}, \citenamefont {Krishna~Kumar}, \citenamefont {Principi}, \citenamefont {Torre}, \citenamefont {Ben~Shalom}, \citenamefont {Taniguchi}, \citenamefont {Watanabe}, \citenamefont {Grigorieva}, \citenamefont {Polini}, \citenamefont {Geim},\ and\ \citenamefont {Bandurin}}]{Berdyugin2019}%
  \BibitemOpen
  \bibfield  {author} {\bibinfo {author} {\bibfnamefont {A.~I.}\ \bibnamefont {Berdyugin}}, \bibinfo {author} {\bibfnamefont {S.~G.}\ \bibnamefont {Xu}}, \bibinfo {author} {\bibfnamefont {F.~M.~D.}\ \bibnamefont {Pellegrino}}, \bibinfo {author} {\bibfnamefont {R.}~\bibnamefont {Krishna~Kumar}}, \bibinfo {author} {\bibfnamefont {A.}~\bibnamefont {Principi}}, \bibinfo {author} {\bibfnamefont {I.}~\bibnamefont {Torre}}, \bibinfo {author} {\bibfnamefont {M.}~\bibnamefont {Ben~Shalom}}, \bibinfo {author} {\bibfnamefont {T.}~\bibnamefont {Taniguchi}}, \bibinfo {author} {\bibfnamefont {K.}~\bibnamefont {Watanabe}}, \bibinfo {author} {\bibfnamefont {I.~V.}\ \bibnamefont {Grigorieva}}, \bibinfo {author} {\bibfnamefont {M.}~\bibnamefont {Polini}}, \bibinfo {author} {\bibfnamefont {A.~K.}\ \bibnamefont {Geim}}, \ and\ \bibinfo {author} {\bibfnamefont {D.~A.}\ \bibnamefont {Bandurin}},\ }\href {\doibase 10.1126/science.aau0685} {\bibfield  {journal} {\bibinfo  {journal} {Science}\ }\textbf {\bibinfo {volume} {364}},\
  \bibinfo {pages} {162} (\bibinfo {year} {2019})}\BibitemShut {NoStop}%
\bibitem [{\citenamefont {Kim}\ \emph {et~al.}(2020)\citenamefont {Kim}, \citenamefont {Xu}, \citenamefont {Berdyugin}, \citenamefont {Principi}, \citenamefont {Slizovskiy}, \citenamefont {Xin}, \citenamefont {Kumaravadivel}, \citenamefont {Kuang}, \citenamefont {Hamer}, \citenamefont {Kumar}, \citenamefont {Gorbachev}, \citenamefont {Watanabe}, \citenamefont {Taniguchi}, \citenamefont {Grigorieva}, \citenamefont {Fal’ko}, \citenamefont {Polini},\ and\ \citenamefont {Geim}}]{Kim2020viscosityscreening}%
  \BibitemOpen
  \bibfield  {author} {\bibinfo {author} {\bibfnamefont {M.}~\bibnamefont {Kim}}, \bibinfo {author} {\bibfnamefont {S.~G.}\ \bibnamefont {Xu}}, \bibinfo {author} {\bibfnamefont {A.~I.}\ \bibnamefont {Berdyugin}}, \bibinfo {author} {\bibfnamefont {A.}~\bibnamefont {Principi}}, \bibinfo {author} {\bibfnamefont {S.}~\bibnamefont {Slizovskiy}}, \bibinfo {author} {\bibfnamefont {N.}~\bibnamefont {Xin}}, \bibinfo {author} {\bibfnamefont {P.}~\bibnamefont {Kumaravadivel}}, \bibinfo {author} {\bibfnamefont {W.}~\bibnamefont {Kuang}}, \bibinfo {author} {\bibfnamefont {M.}~\bibnamefont {Hamer}}, \bibinfo {author} {\bibfnamefont {R.~K.}\ \bibnamefont {Kumar}}, \bibinfo {author} {\bibfnamefont {R.~V.}\ \bibnamefont {Gorbachev}}, \bibinfo {author} {\bibfnamefont {K.}~\bibnamefont {Watanabe}}, \bibinfo {author} {\bibfnamefont {T.}~\bibnamefont {Taniguchi}}, \bibinfo {author} {\bibfnamefont {I.~V.}\ \bibnamefont {Grigorieva}}, \bibinfo {author} {\bibfnamefont {V.~I.}\ \bibnamefont {Fal’ko}}, \bibinfo {author}
  {\bibfnamefont {M.}~\bibnamefont {Polini}}, \ and\ \bibinfo {author} {\bibfnamefont {A.~K.}\ \bibnamefont {Geim}},\ }\href {\doibase https://doi.org/10.1038/s41467-020-15829-1} {\bibfield  {journal} {\bibinfo  {journal} {Nature Communications}\ }\textbf {\bibinfo {volume} {11}},\ \bibinfo {pages} {2339} (\bibinfo {year} {2020})}\BibitemShut {NoStop}%
\bibitem [{\citenamefont {Talanov}\ \emph {et~al.}(2024)\citenamefont {Talanov}, \citenamefont {Waissman}, \citenamefont {Hui}, \citenamefont {Skinner}, \citenamefont {Watanabe}, \citenamefont {Taniguchi},\ and\ \citenamefont {Kim}}]{talanov2024viscousdissipationthermal}%
  \BibitemOpen
  \bibfield  {author} {\bibinfo {author} {\bibfnamefont {A.}~\bibnamefont {Talanov}}, \bibinfo {author} {\bibfnamefont {J.}~\bibnamefont {Waissman}}, \bibinfo {author} {\bibfnamefont {A.}~\bibnamefont {Hui}}, \bibinfo {author} {\bibfnamefont {B.}~\bibnamefont {Skinner}}, \bibinfo {author} {\bibfnamefont {K.}~\bibnamefont {Watanabe}}, \bibinfo {author} {\bibfnamefont {T.}~\bibnamefont {Taniguchi}}, \ and\ \bibinfo {author} {\bibfnamefont {P.}~\bibnamefont {Kim}},\ }\href {https://arxiv.org/abs/2406.13799} {\  (\bibinfo {year} {2024})},\ \Eprint {http://arxiv.org/abs/2406.13799} {arXiv:2406.13799 [cond-mat.mes-hall]} \BibitemShut {NoStop}%
\bibitem [{\citenamefont {Ku}\ \emph {et~al.}(2020)\citenamefont {Ku}, \citenamefont {Zhou}, \citenamefont {Li}, \citenamefont {Shin}, \citenamefont {Shi}, \citenamefont {Burch}, \citenamefont {Zhang}, \citenamefont {Casola}, \citenamefont {Taniguchi}, \citenamefont {Watanabe}, \citenamefont {Kim}, \citenamefont {Yacoby},\ and\ \citenamefont {Walsworth}}]{Ku2020NVhydro}%
  \BibitemOpen
  \bibfield  {author} {\bibinfo {author} {\bibfnamefont {M.~J.}\ \bibnamefont {Ku}}, \bibinfo {author} {\bibfnamefont {T.~X.}\ \bibnamefont {Zhou}}, \bibinfo {author} {\bibfnamefont {Q.}~\bibnamefont {Li}}, \bibinfo {author} {\bibfnamefont {Y.~J.}\ \bibnamefont {Shin}}, \bibinfo {author} {\bibfnamefont {J.~K.}\ \bibnamefont {Shi}}, \bibinfo {author} {\bibfnamefont {C.}~\bibnamefont {Burch}}, \bibinfo {author} {\bibfnamefont {H.}~\bibnamefont {Zhang}}, \bibinfo {author} {\bibfnamefont {F.}~\bibnamefont {Casola}}, \bibinfo {author} {\bibfnamefont {T.}~\bibnamefont {Taniguchi}}, \bibinfo {author} {\bibfnamefont {K.}~\bibnamefont {Watanabe}}, \bibinfo {author} {\bibfnamefont {P.}~\bibnamefont {Kim}}, \bibinfo {author} {\bibfnamefont {A.}~\bibnamefont {Yacoby}}, \ and\ \bibinfo {author} {\bibfnamefont {R.~L.}\ \bibnamefont {Walsworth}},\ }\href {https://doi.org/10.1038/s41586-020-2507-2} {\bibfield  {journal} {\bibinfo  {journal} {Nature}\ }\textbf {\bibinfo {volume} {583}},\ \bibinfo {pages} {537} (\bibinfo
  {year} {2020})}\BibitemShut {NoStop}%
\bibitem [{\citenamefont {Kumar}\ \emph {et~al.}(2022)\citenamefont {Kumar}, \citenamefont {Birkbeck}, \citenamefont {Sulpizio}, \citenamefont {Perello}, \citenamefont {Taniguchi}, \citenamefont {Watanabe}, \citenamefont {Reuven}, \citenamefont {Scaffidi}, \citenamefont {Stern}, \citenamefont {Geim},\ and\ \citenamefont {Ilani}}]{Kumar2022LS}%
  \BibitemOpen
  \bibfield  {author} {\bibinfo {author} {\bibfnamefont {C.}~\bibnamefont {Kumar}}, \bibinfo {author} {\bibfnamefont {J.}~\bibnamefont {Birkbeck}}, \bibinfo {author} {\bibfnamefont {J.~A.}\ \bibnamefont {Sulpizio}}, \bibinfo {author} {\bibfnamefont {D.}~\bibnamefont {Perello}}, \bibinfo {author} {\bibfnamefont {T.}~\bibnamefont {Taniguchi}}, \bibinfo {author} {\bibfnamefont {K.}~\bibnamefont {Watanabe}}, \bibinfo {author} {\bibfnamefont {O.}~\bibnamefont {Reuven}}, \bibinfo {author} {\bibfnamefont {T.}~\bibnamefont {Scaffidi}}, \bibinfo {author} {\bibfnamefont {A.}~\bibnamefont {Stern}}, \bibinfo {author} {\bibfnamefont {A.~K.}\ \bibnamefont {Geim}}, \ and\ \bibinfo {author} {\bibfnamefont {S.}~\bibnamefont {Ilani}},\ }\href {\doibase 10.1038/s41586-022-05002-7} {\bibfield  {journal} {\bibinfo  {journal} {Nature}\ }\textbf {\bibinfo {volume} {609}},\ \bibinfo {pages} {276} (\bibinfo {year} {2022})}\BibitemShut {NoStop}%
\bibitem [{\citenamefont {Sulpizio}\ \emph {et~al.}(2019)\citenamefont {Sulpizio}, \citenamefont {Ella}, \citenamefont {Rozen}, \citenamefont {Birkbeck}, \citenamefont {Perello}, \citenamefont {Dutta}, \citenamefont {Ben-Shalom}, \citenamefont {Taniguchi}, \citenamefont {Watanabe}, \citenamefont {Holder}, \citenamefont {Queiroz}, \citenamefont {Principi}, \citenamefont {Stern}, \citenamefont {Scaffidi}, \citenamefont {Geim},\ and\ \citenamefont {Ilani}}]{sulpizio2019}%
  \BibitemOpen
  \bibfield  {author} {\bibinfo {author} {\bibfnamefont {J.~A.}\ \bibnamefont {Sulpizio}}, \bibinfo {author} {\bibfnamefont {L.}~\bibnamefont {Ella}}, \bibinfo {author} {\bibfnamefont {A.}~\bibnamefont {Rozen}}, \bibinfo {author} {\bibfnamefont {J.}~\bibnamefont {Birkbeck}}, \bibinfo {author} {\bibfnamefont {D.~J.}\ \bibnamefont {Perello}}, \bibinfo {author} {\bibfnamefont {D.}~\bibnamefont {Dutta}}, \bibinfo {author} {\bibfnamefont {M.}~\bibnamefont {Ben-Shalom}}, \bibinfo {author} {\bibfnamefont {T.}~\bibnamefont {Taniguchi}}, \bibinfo {author} {\bibfnamefont {K.}~\bibnamefont {Watanabe}}, \bibinfo {author} {\bibfnamefont {T.}~\bibnamefont {Holder}}, \bibinfo {author} {\bibfnamefont {R.}~\bibnamefont {Queiroz}}, \bibinfo {author} {\bibfnamefont {A.}~\bibnamefont {Principi}}, \bibinfo {author} {\bibfnamefont {A.}~\bibnamefont {Stern}}, \bibinfo {author} {\bibfnamefont {T.}~\bibnamefont {Scaffidi}}, \bibinfo {author} {\bibfnamefont {A.~K.}\ \bibnamefont {Geim}}, \ and\ \bibinfo {author} {\bibfnamefont
  {S.}~\bibnamefont {Ilani}},\ }\href {\doibase 10.1038/s41586-019-1788-9} {\bibfield  {journal} {\bibinfo  {journal} {Nature}\ }\textbf {\bibinfo {volume} {576}},\ \bibinfo {pages} {75} (\bibinfo {year} {2019})}\BibitemShut {NoStop}%
\bibitem [{\citenamefont {Jenkins}\ \emph {et~al.}(2022)\citenamefont {Jenkins}, \citenamefont {Baumann}, \citenamefont {Zhou}, \citenamefont {Meynell}, \citenamefont {Daipeng}, \citenamefont {Watanabe}, \citenamefont {Taniguchi}, \citenamefont {Lucas}, \citenamefont {Young},\ and\ \citenamefont {Bleszynski~Jayich}}]{Jenkins2022NVimaginghydrographene}%
  \BibitemOpen
  \bibfield  {author} {\bibinfo {author} {\bibfnamefont {A.}~\bibnamefont {Jenkins}}, \bibinfo {author} {\bibfnamefont {S.}~\bibnamefont {Baumann}}, \bibinfo {author} {\bibfnamefont {H.}~\bibnamefont {Zhou}}, \bibinfo {author} {\bibfnamefont {S.~A.}\ \bibnamefont {Meynell}}, \bibinfo {author} {\bibfnamefont {Y.}~\bibnamefont {Daipeng}}, \bibinfo {author} {\bibfnamefont {K.}~\bibnamefont {Watanabe}}, \bibinfo {author} {\bibfnamefont {T.}~\bibnamefont {Taniguchi}}, \bibinfo {author} {\bibfnamefont {A.}~\bibnamefont {Lucas}}, \bibinfo {author} {\bibfnamefont {A.~F.}\ \bibnamefont {Young}}, \ and\ \bibinfo {author} {\bibfnamefont {A.~C.}\ \bibnamefont {Bleszynski~Jayich}},\ }\href {\doibase 10.1103/PhysRevLett.129.087701} {\bibfield  {journal} {\bibinfo  {journal} {Phys. Rev. Lett.}\ }\textbf {\bibinfo {volume} {129}},\ \bibinfo {pages} {087701} (\bibinfo {year} {2022})}\BibitemShut {NoStop}%
\bibitem [{\citenamefont {Aharon-Steinberg}\ \emph {et~al.}(2022)\citenamefont {Aharon-Steinberg}, \citenamefont {Völkl}, \citenamefont {Kaplan}, \citenamefont {Pariari}, \citenamefont {Roy}, \citenamefont {Holder}, \citenamefont {Wolf}, \citenamefont {Meltzer}, \citenamefont {Myasoedov}, \citenamefont {Huber}, \citenamefont {Yan}, \citenamefont {Falkovich}, \citenamefont {Levitov}, \citenamefont {Hücker},\ and\ \citenamefont {Zeldov}}]{Aharon-Steinberg2022observationvortice}%
  \BibitemOpen
  \bibfield  {author} {\bibinfo {author} {\bibfnamefont {A.}~\bibnamefont {Aharon-Steinberg}}, \bibinfo {author} {\bibfnamefont {T.}~\bibnamefont {Völkl}}, \bibinfo {author} {\bibfnamefont {A.}~\bibnamefont {Kaplan}}, \bibinfo {author} {\bibfnamefont {A.~K.}\ \bibnamefont {Pariari}}, \bibinfo {author} {\bibfnamefont {I.}~\bibnamefont {Roy}}, \bibinfo {author} {\bibfnamefont {T.}~\bibnamefont {Holder}}, \bibinfo {author} {\bibfnamefont {Y.}~\bibnamefont {Wolf}}, \bibinfo {author} {\bibfnamefont {A.~Y.}\ \bibnamefont {Meltzer}}, \bibinfo {author} {\bibfnamefont {Y.}~\bibnamefont {Myasoedov}}, \bibinfo {author} {\bibfnamefont {M.~E.}\ \bibnamefont {Huber}}, \bibinfo {author} {\bibfnamefont {B.}~\bibnamefont {Yan}}, \bibinfo {author} {\bibfnamefont {G.}~\bibnamefont {Falkovich}}, \bibinfo {author} {\bibfnamefont {L.~S.}\ \bibnamefont {Levitov}}, \bibinfo {author} {\bibfnamefont {M.}~\bibnamefont {Hücker}}, \ and\ \bibinfo {author} {\bibfnamefont {E.}~\bibnamefont {Zeldov}},\ }\href {\doibase
  https://doi.org/10.1038/s41586-022-04794-y} {\bibfield  {journal} {\bibinfo  {journal} {Nature}\ }\textbf {\bibinfo {volume} {607}},\ \bibinfo {pages} {74} (\bibinfo {year} {2022})}\BibitemShut {NoStop}%
\bibitem [{\citenamefont {Vool}\ \emph {et~al.}(2021)\citenamefont {Vool}, \citenamefont {Hamo}, \citenamefont {Varnavides}, \citenamefont {Yaxian~Wang}, \citenamefont {Kumar}, \citenamefont {Dovzhenko}, \citenamefont {Qiu}, \citenamefont {Garcia}, \citenamefont {Pierce}, \citenamefont {Gooth}, \citenamefont {Anikeeva}, \citenamefont {Felser}, \citenamefont {Narang},\ and\ \citenamefont {Yacoby}}]{Vool2021imagingviscous}%
  \BibitemOpen
  \bibfield  {author} {\bibinfo {author} {\bibfnamefont {U.}~\bibnamefont {Vool}}, \bibinfo {author} {\bibfnamefont {A.}~\bibnamefont {Hamo}}, \bibinfo {author} {\bibfnamefont {G.}~\bibnamefont {Varnavides}}, \bibinfo {author} {\bibfnamefont {T.~X.~Z.}\ \bibnamefont {Yaxian~Wang}}, \bibinfo {author} {\bibfnamefont {N.}~\bibnamefont {Kumar}}, \bibinfo {author} {\bibfnamefont {Y.}~\bibnamefont {Dovzhenko}}, \bibinfo {author} {\bibfnamefont {Z.}~\bibnamefont {Qiu}}, \bibinfo {author} {\bibfnamefont {C.~A.~C.}\ \bibnamefont {Garcia}}, \bibinfo {author} {\bibfnamefont {A.~T.}\ \bibnamefont {Pierce}}, \bibinfo {author} {\bibfnamefont {J.}~\bibnamefont {Gooth}}, \bibinfo {author} {\bibfnamefont {P.}~\bibnamefont {Anikeeva}}, \bibinfo {author} {\bibfnamefont {C.}~\bibnamefont {Felser}}, \bibinfo {author} {\bibfnamefont {P.}~\bibnamefont {Narang}}, \ and\ \bibinfo {author} {\bibfnamefont {A.}~\bibnamefont {Yacoby}},\ }\href {\doibase https://doi.org/10.1038/s41567-021-01341-w} {\bibfield  {journal} {\bibinfo
  {journal} {Nature Physics}\ }\textbf {\bibinfo {volume} {17}},\ \bibinfo {pages} {1216} (\bibinfo {year} {2021})}\BibitemShut {NoStop}%
\bibitem [{\citenamefont {Braem}\ \emph {et~al.}(2018)\citenamefont {Braem}, \citenamefont {Pellegrino}, \citenamefont {Principi}, \citenamefont {R\"o\"osli}, \citenamefont {Gold}, \citenamefont {Hennel}, \citenamefont {Koski}, \citenamefont {Berl}, \citenamefont {Dietsche}, \citenamefont {Wegscheider}, \citenamefont {Polini}, \citenamefont {Ihn},\ and\ \citenamefont {Ensslin}}]{Braem2018scanninggateviscous}%
  \BibitemOpen
  \bibfield  {author} {\bibinfo {author} {\bibfnamefont {B.~A.}\ \bibnamefont {Braem}}, \bibinfo {author} {\bibfnamefont {F.~M.~D.}\ \bibnamefont {Pellegrino}}, \bibinfo {author} {\bibfnamefont {A.}~\bibnamefont {Principi}}, \bibinfo {author} {\bibfnamefont {M.}~\bibnamefont {R\"o\"osli}}, \bibinfo {author} {\bibfnamefont {C.}~\bibnamefont {Gold}}, \bibinfo {author} {\bibfnamefont {S.}~\bibnamefont {Hennel}}, \bibinfo {author} {\bibfnamefont {J.~V.}\ \bibnamefont {Koski}}, \bibinfo {author} {\bibfnamefont {M.}~\bibnamefont {Berl}}, \bibinfo {author} {\bibfnamefont {W.}~\bibnamefont {Dietsche}}, \bibinfo {author} {\bibfnamefont {W.}~\bibnamefont {Wegscheider}}, \bibinfo {author} {\bibfnamefont {M.}~\bibnamefont {Polini}}, \bibinfo {author} {\bibfnamefont {T.}~\bibnamefont {Ihn}}, \ and\ \bibinfo {author} {\bibfnamefont {K.}~\bibnamefont {Ensslin}},\ }\href {\doibase 10.1103/PhysRevB.98.241304} {\bibfield  {journal} {\bibinfo  {journal} {Phys. Rev. B}\ }\textbf {\bibinfo {volume} {98}},\ \bibinfo {pages}
  {241304} (\bibinfo {year} {2018})}\BibitemShut {NoStop}%
\bibitem [{\citenamefont {Levchenko}\ \emph {et~al.}(2022)\citenamefont {Levchenko}, \citenamefont {Li},\ and\ \citenamefont {Andreev}}]{levchenko2020}%
  \BibitemOpen
  \bibfield  {author} {\bibinfo {author} {\bibfnamefont {A.}~\bibnamefont {Levchenko}}, \bibinfo {author} {\bibfnamefont {S.}~\bibnamefont {Li}}, \ and\ \bibinfo {author} {\bibfnamefont {A.~V.}\ \bibnamefont {Andreev}},\ }\href {\doibase 10.1103/PhysRevB.106.L201306} {\bibfield  {journal} {\bibinfo  {journal} {Phys. Rev. B}\ }\textbf {\bibinfo {volume} {106}},\ \bibinfo {pages} {L201306} (\bibinfo {year} {2022})}\BibitemShut {NoStop}%
\bibitem [{\citenamefont {Shavit}\ \emph {et~al.}(2019)\citenamefont {Shavit}, \citenamefont {Shytov},\ and\ \citenamefont {Falkovich}}]{shavit2019_corbino}%
  \BibitemOpen
  \bibfield  {author} {\bibinfo {author} {\bibfnamefont {M.}~\bibnamefont {Shavit}}, \bibinfo {author} {\bibfnamefont {A.}~\bibnamefont {Shytov}}, \ and\ \bibinfo {author} {\bibfnamefont {G.}~\bibnamefont {Falkovich}},\ }\href {\doibase 10.1103/PhysRevLett.123.026801} {\bibfield  {journal} {\bibinfo  {journal} {Phys. Rev. Lett.}\ }\textbf {\bibinfo {volume} {123}},\ \bibinfo {pages} {026801} (\bibinfo {year} {2019})}\BibitemShut {NoStop}%
\bibitem [{\citenamefont {Holder}\ \emph {et~al.}(2019)\citenamefont {Holder}, \citenamefont {Queiroz}, \citenamefont {Scaffidi}, \citenamefont {Silberstein}, \citenamefont {Rozen}, \citenamefont {Sulpizio}, \citenamefont {Ella}, \citenamefont {Ilani},\ and\ \citenamefont {Stern}}]{holder2019}%
  \BibitemOpen
  \bibfield  {author} {\bibinfo {author} {\bibfnamefont {T.}~\bibnamefont {Holder}}, \bibinfo {author} {\bibfnamefont {R.}~\bibnamefont {Queiroz}}, \bibinfo {author} {\bibfnamefont {T.}~\bibnamefont {Scaffidi}}, \bibinfo {author} {\bibfnamefont {N.}~\bibnamefont {Silberstein}}, \bibinfo {author} {\bibfnamefont {A.}~\bibnamefont {Rozen}}, \bibinfo {author} {\bibfnamefont {J.~A.}\ \bibnamefont {Sulpizio}}, \bibinfo {author} {\bibfnamefont {L.}~\bibnamefont {Ella}}, \bibinfo {author} {\bibfnamefont {S.}~\bibnamefont {Ilani}}, \ and\ \bibinfo {author} {\bibfnamefont {A.}~\bibnamefont {Stern}},\ }\href {\doibase 10.1103/PhysRevB.100.245305} {\bibfield  {journal} {\bibinfo  {journal} {Phys. Rev. B}\ }\textbf {\bibinfo {volume} {100}},\ \bibinfo {pages} {245305} (\bibinfo {year} {2019})}\BibitemShut {NoStop}%
\bibitem [{\citenamefont {Yudhistira}\ \emph {et~al.}(2023)\citenamefont {Yudhistira}, \citenamefont {Afrose},\ and\ \citenamefont {Adam}}]{yudhistira2023nonmonotonic}%
  \BibitemOpen
  \bibfield  {author} {\bibinfo {author} {\bibfnamefont {I.}~\bibnamefont {Yudhistira}}, \bibinfo {author} {\bibfnamefont {R.}~\bibnamefont {Afrose}}, \ and\ \bibinfo {author} {\bibfnamefont {S.}~\bibnamefont {Adam}},\ }\href@noop {} {\  (\bibinfo {year} {2023})},\ \Eprint {http://arxiv.org/abs/2312.09701} {arXiv:2312.09701 [cond-mat.mes-hall]} \BibitemShut {NoStop}%
\bibitem [{\citenamefont {Zeng}\ \emph {et~al.}(2019)\citenamefont {Zeng}, \citenamefont {Li}, \citenamefont {Dietrich}, \citenamefont {Ghosh}, \citenamefont {Watanabe}, \citenamefont {Taniguchi}, \citenamefont {Hone},\ and\ \citenamefont {Dean}}]{Zeng2019}%
  \BibitemOpen
  \bibfield  {author} {\bibinfo {author} {\bibfnamefont {Y.}~\bibnamefont {Zeng}}, \bibinfo {author} {\bibfnamefont {J.~I.~A.}\ \bibnamefont {Li}}, \bibinfo {author} {\bibfnamefont {S.~A.}\ \bibnamefont {Dietrich}}, \bibinfo {author} {\bibfnamefont {O.~M.}\ \bibnamefont {Ghosh}}, \bibinfo {author} {\bibfnamefont {K.}~\bibnamefont {Watanabe}}, \bibinfo {author} {\bibfnamefont {T.}~\bibnamefont {Taniguchi}}, \bibinfo {author} {\bibfnamefont {J.}~\bibnamefont {Hone}}, \ and\ \bibinfo {author} {\bibfnamefont {C.~R.}\ \bibnamefont {Dean}},\ }\href {\doibase 10.1103/PhysRevLett.122.137701} {\bibfield  {journal} {\bibinfo  {journal} {Phys. Rev. Lett.}\ }\textbf {\bibinfo {volume} {122}},\ \bibinfo {pages} {137701} (\bibinfo {year} {2019})}\BibitemShut {NoStop}%
\bibitem [{\citenamefont {Weiss}\ and\ \citenamefont {Welker.}(1954)}]{Weiss1954corbino}%
  \BibitemOpen
  \bibfield  {author} {\bibinfo {author} {\bibfnamefont {H.}~\bibnamefont {Weiss}}\ and\ \bibinfo {author} {\bibfnamefont {H.}~\bibnamefont {Welker.}},\ }\href {\doibase https://doi.org/10.1007/BF01340677} {\bibfield  {journal} {\bibinfo  {journal} {Z. Physik}\ }\textbf {\bibinfo {volume} {138}},\ \bibinfo {pages} {322} (\bibinfo {year} {1954})}\BibitemShut {NoStop}%
\bibitem [{\citenamefont {Blood}\ and\ \citenamefont {Tree}(1971)}]{Blood1971corbino}%
  \BibitemOpen
  \bibfield  {author} {\bibinfo {author} {\bibfnamefont {P.}~\bibnamefont {Blood}}\ and\ \bibinfo {author} {\bibfnamefont {R.~J.}\ \bibnamefont {Tree}},\ }\href {\doibase 10.1088/0022-3727/4/9/101} {\bibfield  {journal} {\bibinfo  {journal} {Journal of Physics D: Applied Physics}\ }\textbf {\bibinfo {volume} {4}},\ \bibinfo {pages} {L29} (\bibinfo {year} {1971})}\BibitemShut {NoStop}%
\bibitem [{\citenamefont {Wieder}(1969)}]{wieder1969corbino}%
  \BibitemOpen
  \bibfield  {author} {\bibinfo {author} {\bibfnamefont {H.~H.}\ \bibnamefont {Wieder}},\ }\href {\doibase 10.1063/1.1658182} {\bibfield  {journal} {\bibinfo  {journal} {Journal of Applied Physics}\ }\textbf {\bibinfo {volume} {40}},\ \bibinfo {pages} {3320} (\bibinfo {year} {1969})}\BibitemShut {NoStop}%
\bibitem [{\citenamefont {Alekseev}(2016)}]{Alekseev2016}%
  \BibitemOpen
  \bibfield  {author} {\bibinfo {author} {\bibfnamefont {P.~S.}\ \bibnamefont {Alekseev}},\ }\href {\doibase 10.1103/PhysRevLett.117.166601} {\bibfield  {journal} {\bibinfo  {journal} {Phys. Rev. Lett.}\ }\textbf {\bibinfo {volume} {117}},\ \bibinfo {pages} {166601} (\bibinfo {year} {2016})}\BibitemShut {NoStop}%
\bibitem [{\citenamefont {Dean}\ \emph {et~al.}(2010)\citenamefont {Dean}, \citenamefont {Young}, \citenamefont {Meric}, \citenamefont {Lee}, \citenamefont {Wang}, \citenamefont {Sorgenfrei}, \citenamefont {Watanabe}, \citenamefont {Taniguchi}, \citenamefont {Kim}, \citenamefont {Shepard},\ and\ \citenamefont {Hone}}]{Dean10}%
  \BibitemOpen
  \bibfield  {author} {\bibinfo {author} {\bibfnamefont {C.~R.}\ \bibnamefont {Dean}}, \bibinfo {author} {\bibfnamefont {A.~F.}\ \bibnamefont {Young}}, \bibinfo {author} {\bibfnamefont {I.}~\bibnamefont {Meric}}, \bibinfo {author} {\bibfnamefont {C.}~\bibnamefont {Lee}}, \bibinfo {author} {\bibfnamefont {L.}~\bibnamefont {Wang}}, \bibinfo {author} {\bibfnamefont {S.}~\bibnamefont {Sorgenfrei}}, \bibinfo {author} {\bibfnamefont {K.}~\bibnamefont {Watanabe}}, \bibinfo {author} {\bibfnamefont {T.}~\bibnamefont {Taniguchi}}, \bibinfo {author} {\bibfnamefont {P.}~\bibnamefont {Kim}}, \bibinfo {author} {\bibfnamefont {K.~L.}\ \bibnamefont {Shepard}}, \ and\ \bibinfo {author} {\bibfnamefont {J.}~\bibnamefont {Hone}},\ }\href {\doibase 10.1038/nnano.2010.172} {\bibfield  {journal} {\bibinfo  {journal} {Nature Nanotechnology}\ }\textbf {\bibinfo {volume} {5}},\ \bibinfo {pages} {722–} (\bibinfo {year} {2010})}\BibitemShut {NoStop}%
\bibitem [{\citenamefont {Chen}\ \emph {et~al.}(2008)\citenamefont {Chen}, \citenamefont {Jang}, \citenamefont {Xiao}, \citenamefont {Ishigami},\ and\ \citenamefont {Fuhrer}}]{chen2008graphene_scattering}%
  \BibitemOpen
  \bibfield  {author} {\bibinfo {author} {\bibfnamefont {J.-H.}\ \bibnamefont {Chen}}, \bibinfo {author} {\bibfnamefont {C.}~\bibnamefont {Jang}}, \bibinfo {author} {\bibfnamefont {S.}~\bibnamefont {Xiao}}, \bibinfo {author} {\bibfnamefont {M.}~\bibnamefont {Ishigami}}, \ and\ \bibinfo {author} {\bibfnamefont {M.~S.}\ \bibnamefont {Fuhrer}},\ }\href {\doibase 10.1038/nnano.2008.58} {\bibfield  {journal} {\bibinfo  {journal} {Nature Nanotechnology}\ }\textbf {\bibinfo {volume} {3}},\ \bibinfo {pages} {206} (\bibinfo {year} {2008})}\BibitemShut {NoStop}%
\bibitem [{\citenamefont {Li}\ \emph {et~al.}(2016)\citenamefont {Li}, \citenamefont {Tan}, \citenamefont {Zou}, \citenamefont {Stabile}, \citenamefont {Seiwell}, \citenamefont {Watanabe}, \citenamefont {Taniguchi}, \citenamefont {Louie},\ and\ \citenamefont {Zhu}}]{li2016_blgmass}%
  \BibitemOpen
  \bibfield  {author} {\bibinfo {author} {\bibfnamefont {J.}~\bibnamefont {Li}}, \bibinfo {author} {\bibfnamefont {L.~Z.}\ \bibnamefont {Tan}}, \bibinfo {author} {\bibfnamefont {K.}~\bibnamefont {Zou}}, \bibinfo {author} {\bibfnamefont {A.~A.}\ \bibnamefont {Stabile}}, \bibinfo {author} {\bibfnamefont {D.~J.}\ \bibnamefont {Seiwell}}, \bibinfo {author} {\bibfnamefont {K.}~\bibnamefont {Watanabe}}, \bibinfo {author} {\bibfnamefont {T.}~\bibnamefont {Taniguchi}}, \bibinfo {author} {\bibfnamefont {S.~G.}\ \bibnamefont {Louie}}, \ and\ \bibinfo {author} {\bibfnamefont {J.}~\bibnamefont {Zhu}},\ }\href {\doibase 10.1103/PhysRevB.94.161406} {\bibfield  {journal} {\bibinfo  {journal} {Phys. Rev. B}\ }\textbf {\bibinfo {volume} {94}},\ \bibinfo {pages} {161406} (\bibinfo {year} {2016})}\BibitemShut {NoStop}%
\bibitem [{\citenamefont {Principi}\ \emph {et~al.}(2016)\citenamefont {Principi}, \citenamefont {Vignale}, \citenamefont {Carrega},\ and\ \citenamefont {Polini}}]{principi2016}%
  \BibitemOpen
  \bibfield  {author} {\bibinfo {author} {\bibfnamefont {A.}~\bibnamefont {Principi}}, \bibinfo {author} {\bibfnamefont {G.}~\bibnamefont {Vignale}}, \bibinfo {author} {\bibfnamefont {M.}~\bibnamefont {Carrega}}, \ and\ \bibinfo {author} {\bibfnamefont {M.}~\bibnamefont {Polini}},\ }\href {\doibase 10.1103/PhysRevB.93.125410} {\bibfield  {journal} {\bibinfo  {journal} {Phys. Rev. B}\ }\textbf {\bibinfo {volume} {93}},\ \bibinfo {pages} {125410} (\bibinfo {year} {2016})}\BibitemShut {NoStop}%
\bibitem [{\citenamefont {Kovtun}\ \emph {et~al.}(2005)\citenamefont {Kovtun}, \citenamefont {Son},\ and\ \citenamefont {Starinets}}]{kovtun2005viscositybound}%
  \BibitemOpen
  \bibfield  {author} {\bibinfo {author} {\bibfnamefont {P.~K.}\ \bibnamefont {Kovtun}}, \bibinfo {author} {\bibfnamefont {D.~T.}\ \bibnamefont {Son}}, \ and\ \bibinfo {author} {\bibfnamefont {A.~O.}\ \bibnamefont {Starinets}},\ }\href {\doibase 10.1103/PhysRevLett.94.111601} {\bibfield  {journal} {\bibinfo  {journal} {Phys. Rev. Lett.}\ }\textbf {\bibinfo {volume} {94}},\ \bibinfo {pages} {111601} (\bibinfo {year} {2005})}\BibitemShut {NoStop}%
\bibitem [{\citenamefont {Son}(2007)}]{son2007prl}%
  \BibitemOpen
  \bibfield  {author} {\bibinfo {author} {\bibfnamefont {D.~T.}\ \bibnamefont {Son}},\ }\href {\doibase 10.1103/PhysRevLett.98.020604} {\bibfield  {journal} {\bibinfo  {journal} {Phys. Rev. Lett.}\ }\textbf {\bibinfo {volume} {98}},\ \bibinfo {pages} {020604} (\bibinfo {year} {2007})}\BibitemShut {NoStop}%
\bibitem [{\citenamefont {Son}\ and\ \citenamefont {Starinets}(2007)}]{son2007arnpp}%
  \BibitemOpen
  \bibfield  {author} {\bibinfo {author} {\bibfnamefont {D.~T.}\ \bibnamefont {Son}}\ and\ \bibinfo {author} {\bibfnamefont {A.~O.}\ \bibnamefont {Starinets}},\ }\href {\doibase 10.1146/annurev.nucl.57.090506.123120} {\bibfield  {journal} {\bibinfo  {journal} {Annual Review of Nuclear and Particle Science}\ }\textbf {\bibinfo {volume} {57}},\ \bibinfo {pages} {95} (\bibinfo {year} {2007})}\BibitemShut {NoStop}%
\bibitem [{\citenamefont {Karsch}\ \emph {et~al.}(2008)\citenamefont {Karsch}, \citenamefont {Kharzeev},\ and\ \citenamefont {Tuchin}}]{KARSCH2008}%
  \BibitemOpen
  \bibfield  {author} {\bibinfo {author} {\bibfnamefont {F.}~\bibnamefont {Karsch}}, \bibinfo {author} {\bibfnamefont {D.}~\bibnamefont {Kharzeev}}, \ and\ \bibinfo {author} {\bibfnamefont {K.}~\bibnamefont {Tuchin}},\ }\href {\doibase https://doi.org/10.1016/j.physletb.2008.01.080} {\bibfield  {journal} {\bibinfo  {journal} {Physics Letters B}\ }\textbf {\bibinfo {volume} {663}},\ \bibinfo {pages} {217 } (\bibinfo {year} {2008})}\BibitemShut {NoStop}%
\bibitem [{\citenamefont {Sachdev}\ and\ \citenamefont {Müller}(2009)}]{Sachdev2009_JPCM}%
  \BibitemOpen
  \bibfield  {author} {\bibinfo {author} {\bibfnamefont {S.}~\bibnamefont {Sachdev}}\ and\ \bibinfo {author} {\bibfnamefont {M.}~\bibnamefont {Müller}},\ }\href {\doibase 10.1088/0953-8984/21/16/164216} {\bibfield  {journal} {\bibinfo  {journal} {Journal of Physics: Condensed Matter}\ }\textbf {\bibinfo {volume} {21}},\ \bibinfo {pages} {164216} (\bibinfo {year} {2009})}\BibitemShut {NoStop}%
\bibitem [{\citenamefont {Crossno}\ \emph {et~al.}(2016)\citenamefont {Crossno}, \citenamefont {Shi}, \citenamefont {Wang}, \citenamefont {Liu}, \citenamefont {Harzheim}, \citenamefont {Lucas}, \citenamefont {Sachdev}, \citenamefont {Kim}, \citenamefont {Taniguchi}, \citenamefont {Watanabe}, \citenamefont {Ohki},\ and\ \citenamefont {Fong}}]{Crossno2016}%
  \BibitemOpen
  \bibfield  {author} {\bibinfo {author} {\bibfnamefont {J.}~\bibnamefont {Crossno}}, \bibinfo {author} {\bibfnamefont {J.~K.}\ \bibnamefont {Shi}}, \bibinfo {author} {\bibfnamefont {K.}~\bibnamefont {Wang}}, \bibinfo {author} {\bibfnamefont {X.}~\bibnamefont {Liu}}, \bibinfo {author} {\bibfnamefont {A.}~\bibnamefont {Harzheim}}, \bibinfo {author} {\bibfnamefont {A.}~\bibnamefont {Lucas}}, \bibinfo {author} {\bibfnamefont {S.}~\bibnamefont {Sachdev}}, \bibinfo {author} {\bibfnamefont {P.}~\bibnamefont {Kim}}, \bibinfo {author} {\bibfnamefont {T.}~\bibnamefont {Taniguchi}}, \bibinfo {author} {\bibfnamefont {K.}~\bibnamefont {Watanabe}}, \bibinfo {author} {\bibfnamefont {T.~A.}\ \bibnamefont {Ohki}}, \ and\ \bibinfo {author} {\bibfnamefont {K.~C.}\ \bibnamefont {Fong}},\ }\href {\doibase 10.1126/science.aad0343} {\bibfield  {journal} {\bibinfo  {journal} {Science}\ }\textbf {\bibinfo {volume} {351}},\ \bibinfo {pages} {1058} (\bibinfo {year} {2016})}\BibitemShut {NoStop}%
\bibitem [{\citenamefont {Wang}\ \emph {et~al.}(2013)\citenamefont {Wang}, \citenamefont {Meric}, \citenamefont {Huang}, \citenamefont {Gao}, \citenamefont {Gao}, \citenamefont {Tran}, \citenamefont {Taniguchi}, \citenamefont {Watanabe}, \citenamefont {Campos}, \citenamefont {Muller}, \citenamefont {Guo}, \citenamefont {Kim}, \citenamefont {Hone}, \citenamefont {Shepard},\ and\ \citenamefont {Dean}}]{Wang2013}%
  \BibitemOpen
  \bibfield  {author} {\bibinfo {author} {\bibfnamefont {L.}~\bibnamefont {Wang}}, \bibinfo {author} {\bibfnamefont {I.}~\bibnamefont {Meric}}, \bibinfo {author} {\bibfnamefont {P.}~\bibnamefont {Huang}}, \bibinfo {author} {\bibfnamefont {Q.}~\bibnamefont {Gao}}, \bibinfo {author} {\bibfnamefont {Y.}~\bibnamefont {Gao}}, \bibinfo {author} {\bibfnamefont {H.}~\bibnamefont {Tran}}, \bibinfo {author} {\bibfnamefont {T.}~\bibnamefont {Taniguchi}}, \bibinfo {author} {\bibfnamefont {K.}~\bibnamefont {Watanabe}}, \bibinfo {author} {\bibfnamefont {L.}~\bibnamefont {Campos}}, \bibinfo {author} {\bibfnamefont {D.}~\bibnamefont {Muller}}, \bibinfo {author} {\bibfnamefont {J.}~\bibnamefont {Guo}}, \bibinfo {author} {\bibfnamefont {P.}~\bibnamefont {Kim}}, \bibinfo {author} {\bibfnamefont {J.}~\bibnamefont {Hone}}, \bibinfo {author} {\bibfnamefont {K.~L.}\ \bibnamefont {Shepard}}, \ and\ \bibinfo {author} {\bibfnamefont {C.~R.}\ \bibnamefont {Dean}},\ }\href@noop {} {\bibfield  {journal} {\bibinfo  {journal} {Science}\
  }\textbf {\bibinfo {volume} {342}},\ \bibinfo {pages} {614} (\bibinfo {year} {2013})}\BibitemShut {NoStop}%
\end{thebibliography}%

\newpage

\end{document}